\documentclass[draftcls,onecolumn, 12pt]{IEEEtran}

\usepackage[cmex10]{amsmath}
\usepackage{amssymb,amsfonts,mathrsfs}
\usepackage{cite}
\usepackage{array}
\usepackage{enumerate}
\usepackage[dvips]{graphicx}
\graphicspath{{./Figures/}}
\usepackage{psfrag}
\usepackage{url}
\usepackage[dvips]{hyperref}

\usepackage{color}

\newtheorem{Proposition}{Proposition}
\newtheorem{Lemma}{Lemma}

\newtheorem{Theorem}{Theorem}
\newtheorem{Example}{Example}

\newtheorem{Assumption}{Assumption}

\newtheorem{Proposition_A}{Proposition}[section]

\newcommand{\Real}{\mathbb{R}}

\newcommand{\ud}{\mathrm{d}}
\newcommand{\E}{\mathbb{E}}
\newcommand{\var}{{\textrm{var}}}
\renewcommand{\P}{\mathbb{P}}

\newcommand{\ofrac}[1]{{\frac{1}{#1}}}
\newcommand{\ddfrac}[2]{{\frac{\ud #1}{\ud #2}}}

\newcommand{\Lh}[1]{\ell_{#1}}
\newcommand{\LLh}[1]{\log{\Lh{#1}}}
\newcommand{\LLp}[2]{\log^{#2}{\Lh{#1}}}
\newcommand{\LLn}[1]{\mathcal{L}_{#1}^{(n)}}
\newcommand{\LLw}[2]{\mathcal{L}_{#1}^{#2}}

\newcommand{\Bigmid}{{\ \Big| \ }}
\newcommand{\indicator}[1]{{\bf 1}_{\{#1\}}}
\newcommand{\EE}[2]{{\mathbb{E}_{#1}\left[{#2}\right]}}
\newcommand{\Lambdas}[2]{{\Lambda^*_{#1}\left({#2}\right)}}

\newcommand{\hide}[1]{}

\newcommand{\hhide}[1]{}

\begin{document}

\title{The Value of Feedback in Decentralized Detection
\thanks{
This research was supported by the MOE AcRF Tier 1 Grant M52040000. Preliminary versions of parts of this paper were presented at the International Symposium on Wireless and Pervasive Computing, Feb 2011, and at the IEEE International Conference on Acoustics, Speech and Signal Processing, May 2011.  The author is with the Nanyang Technological university, Singapore. E-mail: \texttt{wptay@ntu.edu.sg}
}
}

\author{Wee~Peng~Tay,~\IEEEmembership{Member,~IEEE}
       }


\maketitle \thispagestyle{empty}


\begin{abstract}
We consider the decentralized binary hypothesis testing problem in networks with feedback, where some or all of the sensors have access to compressed summaries of other sensors' observations. We study certain two-message feedback architectures, in which every sensor sends two messages to a fusion center, with the second message based on full or partial knowledge of the first messages of the other sensors. We also study one-message feedback architectures, in which each sensor sends one message to a fusion center, with a group of sensors having full or partial knowledge of the messages from the sensors not in that group. Under either a Neyman-Pearson or a Bayesian formulation, we show that the asymptotically  optimal (in the limit of a large number of sensors) detection performance (as quantified by error exponents) does not benefit from the feedback messages, if the fusion center remembers all sensor messages. However, feedback can improve the Bayesian detection performance in the one-message feedback architecture if the fusion center has limited memory; for that case, we determine the corresponding optimal error exponents. 
\end{abstract}


\section{Introduction}\label{sect:Introduction}

In the problem of decentralized detection, introduced by Tenney and Sandell \cite{TenSan:81}, each one of several sensors makes an observation and sends a summary by first applying a quantization function to its observation and then communicating the result to a fusion center. The fusion center makes a final decision based on all of the sensor messages. The goal is to design the sensor quantization functions and the fusion rule so as to minimize a cost function, such as the probability of an incorrect final decision.

In this paper we consider sensor network architectures that are more complex than those in \cite{TenSan:81}, and which involve feedback: some or all of the sensors have access to compressed summaries of other sensors' observations. We are interested in characterizing the performance under different architectures, and, in particular, to determine whether the presence of feedback can substantially enhance performance. Because an exact analysis  is seemingly intractable, we focus on the asymptotic regime, involving a large number of sensors, and quantify performance in terms of error exponents. The somewhat unexpected conclusion is that for most of the models considered in this paper,  feedback does not improve performance in binary hypothesis testing. The only exception we have found is Bayesian hypothesis testing in a ``daisy-chain architecture'' (cf.\ Section \ref{sect:ProblemFormulation}) where the fusion center has limited memory. In this configuration, feedback can result in a better optimal error exponent. 

\subsection{Related Literature}

The decentralized detection problem has been widely studied for various network architectures, including the above described ``parallel'' configuration of \cite{TenSan:81} (see \cite{ChaVar:86,PolTsi:90,WilWar:92,Tsi:93a,Tsi:93,IrvTsi:94,VisVar:97,CheVar:02,ChaVee:03,CheWil:05,Kas:06,LiuChe:06}), tandem networks \cite{VisThoTum:88,TanPatKle:91,PapAth:92,TayTsiWin:J08c}, and bounded height tree architectures\cite{EkcTen:82,ReiNol:90a,TanPatKle:93,PetPatKle:94,AlhVar:95,LinCheVar:05,TayTsiWin:J08a,TayTsiWin:J08b,TayTsiWin:J09a}. For sensor observations not conditionally independent given the hypothesis, the problem of designing the quantization functions is known to be NP-hard \cite{TsiAth:85}. For this reason, most of the literature assumes that the sensor observations are conditionally independent. (Some works \cite{ChaVee:06,MisTon:08,SunPooYu:09} have considered the case of correlated observations under Gaussian models, but without addressing the problem of designing optimal quantization functions.) 

Non-tree networks are harder to analyze because the different messages received by a sensor are not in general conditionally independent. While some structural properties of optimal decision rules are available (see, e.g., \cite{KreWil:10}), not much is known about the optimal performance. Networks with feedback face the same difficulty, and the relevant literature (discussed in the next paragraph) is limited.

A variety of feedback architectures, under a Bayesian formulation, have been studied 
in \cite{AlhVar:95,AlhVar:96}. These references 
show that it is person-by-person optimal for every sensor to use a likelihood ratio quantizer, with thresholds that depend on the feedback messages. However, because of the difficulty of optimizing these thresholds when the number of sensors becomes large, it is difficult to analytically compare the performance of networks with and without feedback. Numerical examples in \cite{AlhVar:96} show that a system with feedback has lower probability of error, as expected. To better understand the asymptotics of the error probability, \cite{ShaPap:93} studies the error probability decay rate under a Neyman-Pearson formulation for two different feedback architectures. For either case, it shows that if the fusion center also has access to the fed back messages, then feedback does not improve the optimal error exponent. References \cite{Zou:Msc09,KreTsiZou:09} consider the Neyman-Pearson problem in a daisy-chain architecture (see Figure \ref{fig:daisychain}), and obtain a similar result. However, the analogous questions under a Bayesian formulation were left open in \cite{ShaPap:93,Zou:Msc09,KreTsiZou:09}.

\subsection{Summary and Contributions}
In this paper, we 
revisit some of the architectures studied in
\cite{ShaPap:93,Zou:Msc09,KreTsiZou:09}, and extend the available results. We also study certain feedback architectures that have not been studied before. In what follows, we describe briefly the architectures that we consider, and summarize our results.

\begin{enumerate}

\item We study a new {\bf two-message sequential feedback} architecture.  Sensors are indexed, and the second message of a sensor can take into account the first message of all sensors with lower indices.  We show that under either the Neyman-Pearson or Bayesian formulation, feedback does not improve the error exponent.

\item 
We consider the {\bf two-message full feedback} architecture studied
in \cite{ShaPap:93}. Here,  each sensor gets to transmit two messages, and the second message can take into account the first messages of all sensors. We resolve an open problem for the Bayesian formulation, by showing that there is no performance gain over the non-feedback case. We also provide a variant of the result of \cite{ShaPap:93} for the Neyman-Pearson case. Our model is somewhat more general than that in \cite{ShaPap:93}, because we do not restrict the sensors' raw observations and the sensor messages to be finitely-valued. More crucially, we also remove the constraint in \cite{ShaPap:93} that the feedback message alphabet can grow at most subexponentially with the number of sensors.

\item We consider the {\bf one-message sequential feedback} architecture studied in \cite{SmiSor:00,AceDahLobOzd:11} (under the name of ``full observation network topology''), where sensors are indexed, and each sensor knows the messages of all sensors with lower indices. Unlike \cite{SmiSor:00,AceDahLobOzd:11}, which investigate ``myopic'' strategies where each sensor selfishly minimizes its local error probability, we show that if there is cooperation amongst sensors so that the last sensor makes the final decision for the whole network, there is no loss of asymptotic optimality if sensors other than the last ignore information from the other sensors, for both the Neyman-Pearson and the Bayesian formulation.

\item 
We consider the {\bf daisy chain} or {\bf one-message} architectures studied in \cite{Zou:Msc09}, under which the sensors are divided into two groups, and sensors in the second group have full or partial knowledge of the messages sent by the first group. Reference \cite{Zou:Msc09} dealt with the Neyman-Pearson formulation. In this paper, we turn to the Bayesian formulation and resolve several questions that had been left open. 
\begin{enumerate}
\item 
In a {\bf full feedback daisy chain}, sensors in the second group, as well as the fusion center,
have access to all messages sent by sensors in the first group. Similar to the Neyman-Pearson case, we show that the Bayesian optimal error exponent is the same as for a parallel configuration with the same number of sensors; in particular, feedback offers no performance improvement.

\item
In a {\bf  restricted feedback daisy chain}, the second group of sensors, as well as the fusion center, have access to only a 1-bit summary of the messages sent by sensors in the first group. For the Neyman-Pearson formulation, \cite{KreTsiZou:09} shows that feedback does not improve the error exponent. In contrast, for the Bayesian formulation, we show that in general, feeding this 1-bit summary to the second group of sensors can improve the detection performance. We provide sufficient conditions for 
feedback to result in no performance gain. Furthermore, we show that this architecture is strictly inferior to the full feedback daisy chain and the parallel configuration. We also provide a characterization of the optimal error exponent. 
\end{enumerate}
\end{enumerate}

The remainder of the paper is organized as follows.
In Section \ref{sect:ProblemFormulation} we define the model, formulate the problems that we will be studying, and provide some background material.  In Section \ref{sect:TwoMessage}, we study two-message feedback architectures (sequential and full feedback). In Section \ref{sect:OneMessage}, we study one-message feedback architectures. We offer concluding remarks and discuss open problems in Section \ref{sect:Conclusion}. Some mathematical results that we use frequently are presented in the Appendix.

\section{Problem Formulation}\label{sect:ProblemFormulation}

In this section, we describe the feedback architectures of interest, define our model, and present some preliminary results. We consider a decentralized binary detection problem involving $n$ sensors and a fusion center. Each sensor $k$ observes a random variable $X_k$ taking values in some measurable space $(\mathcal{X},\mathcal{F})$, and is distributed according to a measure $\P_j$ under hypothesis $H_j$, for $j=0,1$. Under either hypothesis $H_j$, $j=0,1$, the random variables $X_k$ are assumed to be i.i.d. We use $\E_j$ to denote the expectation operator with respect to (w.r.t.)\ $\P_j$, and $X_1^n$ to denote the vector $(X_1,\ldots,X_n)$. A similar notation, e.g., $Y_1^n$ will be used for other vectors of random variables as well.

Let $\mathcal{T}$ be the set from which messages take their values. In most engineering applications, $\mathcal{T}$ is assumed to be a finite alphabet, although we do not require this restriction. This allows us to model the received messages at the fusion center over noisy channels. Furthermore, we use $\Gamma$ to denote the set of allowed quantization functions, that is, functions $\gamma : \mathcal{X} \mapsto \mathcal{T}$, that can be used to map observations to messages. One possible choice is to let $\Gamma$ consist of all measurable functions. Alternatively, for the problems considered in this paper, it is known that for $\mathcal{T}$ finite, there is no loss of optimality if we let $\Gamma$ be the set of likelihood-ratio quantizers \cite{AlhVar:95,AlhVar:96,KreWil:10}. 

We consider two classes of feedback architectures: the two-message and one-message architectures. 

\subsection{Two-Message Feedback Architectures}\label{subsect:Architectures}

\begin{figure}[!htb]
\begin{center}
\includegraphics[scale=1]{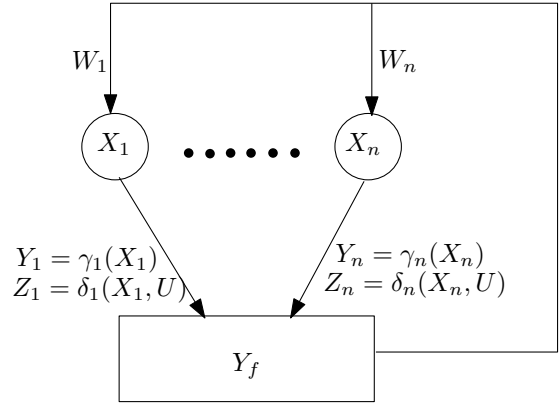}
\caption{A two-message architecture.
}\label{fig:2-message}
\end{center}
\end{figure}

In two-message feedback architectures (see Figure \ref{fig:2-message}), each sensor $k$ sends a message $Y_k = \gamma_k(X_k)$, with $\gamma_k\in\Gamma$, which is a ``quantized'' version of its observation $X_k$, to the fusion center.

We assume that the sensors are indexed in the order that they send their messages to the fusion center.
We consider three forms of feedback under the two-message architecture. 
\begin{itemize}
\item[(a)] {\bf Sequential feedback.} Here, for $k=2,\ldots,n$, 
 the feedback message sent by the fusion center to sensor $k$ is
$W_k = (Y_1,\ldots,Y_{k-1})$, the vector of messages generated by the previous sensors. 
\item[(b)] {\bf Full feedback.} The feedback message sent by the fusion center to sensor $k$ is
the vector $W_k = (Y_1^{k-1},Y_{k+1}^n)$ of messages generated by all of the other sensors.
\item[(c)] {\bf Restricted feedback.} The feedback message sent by the fusion center to sensor $k$ is a function $W_k = f_k(Y_1^{k-1},Y_{k+1}^n)$  of the other sensors' first messages, whose alphabet does not increase with the number of sensors.
\end{itemize}

In all of the above scenarios, each sensor forms a new, second message $Z_k = \delta_k(X_k, W_k)$ based on the additional information $W_k$, and sends it to the fusion center. 

For simplicity, we assume that $Z_k$ takes values in the same alphabet $\mathcal{T}$ and, furthermore, that for any $w$, the function $\delta_k^w(\cdot) = \delta_k(\cdot, w)$ is constrained to belong to the same set  $\Gamma$ that applies to the first round. As alluded to earlier, when $\cal T$ is finite, it is known that there is no loss of optimality if we restrict to log-likelihood ratio quantizers of $X_k$, with thresholds that depend on the received messages. 

Finally, the fusion center makes a decision $Y_f = \gamma_f(Y_1^n,Z_1^n)$. Here, we assume that the fusion center always remembers the first messages $Y_1,\ldots,Y_n$. The collection $(\gamma_f, \gamma_1,\ldots,\gamma_n,\delta_1,\ldots,\delta_n)$ is called a strategy. A sequence of strategies, one for each value of $n$, is called a strategy sequence. We wish to design strategy sequences that that are asymptotically optimal (in the sense of error exponents), as $n$ increases to infinity.

\subsection{One-Message Feedback Architectures}

In one-message architectures, every sensor sends a single message to an intermediate aggregator or the  fusion center, but some of the sensors have access to the messages of some other sensors. Specifically, we consider a one-message sequential feedback architecture\cite{SmiSor:00,AceDahLobOzd:11}, and a daisy chain architecture\cite{Zou:Msc09,KreTsiZou:09}. As before, we let $\Gamma$ be the set of allowed quantization functions.

\begin{itemize}
\item[(a)] {\bf One-message sequential feedback.}  Here, sensor $k$ has access to the messages $Y_1,\ldots,Y_{k-1}$ of all sensors with lower indices. Sensor $k$ forms a message $Y_k = \gamma_k(X_k,Y_1^{k-1})$, and broadcasts it to all sensors with higher indices. The last sensor, $n$, makes a final decision and plays the role of a fusion center. We assume that for any $Y_1^{k-1}$, the mapping from $X_k$ to $Y_k$ belongs to $\Gamma$.
\item[(b)]  {\bf Daisy chain.} This architecture consists of two stages (see Figure \ref{fig:daisychain}) with the first stage involving $m$ sensors and the second $n-m$. Each sensor $k$ in the first stage sends a message $Y_k = \gamma_k(X_k)$ to an aggregator, with $\gamma_k\in\Gamma$. The aggregator forms a message $U$ that is broadcast to all sensors in the second stage and to the fusion center. Each sensor $l$ in the second stage forms a message $Z_l = \delta_l^U(X_l) = \delta_l(X_l, U)$, which depends on its own observation and the message $U$. Again, we assume that $\delta_i^u\in\Gamma$, for every possible value $u$ of $U$. The fusion center makes a final decision using a fusion rule $Y_f = \gamma_f(U,Z_{m+1},\ldots,Z_n)$. We can view the daisy chain as a parallel configuration, in which the fusion center feeds sensors $m+1,\ldots,n$ with a message based on information from sensors $1,\ldots,m$.

\begin{figure}[!htb]
\begin{center}
\includegraphics[scale=1]{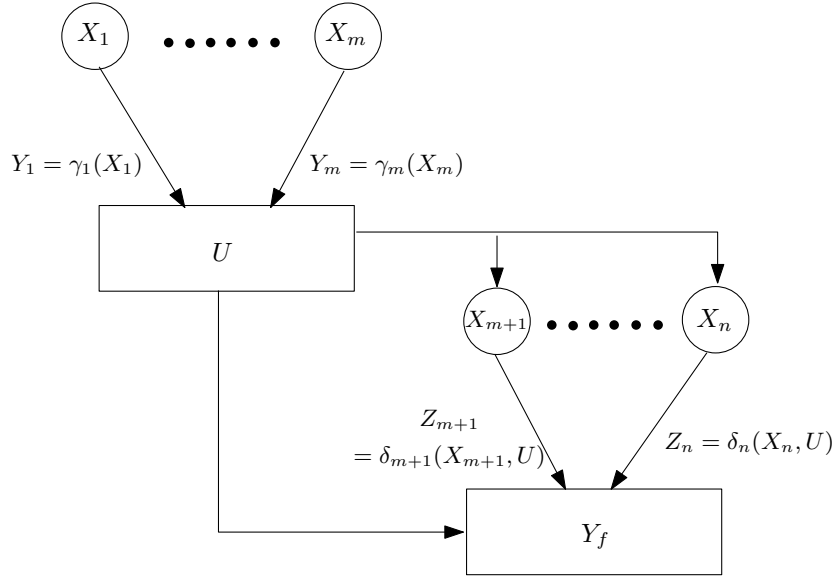}
\caption{The daisy chain architecture.
}\label{fig:daisychain}
\end{center}
\end{figure}

We consider two cases for how $U$ is formed. 
\begin{itemize}
\item[(i)] {\bf Full feedback daisy chain.} 
Here, we let $U = (Y_1,\ldots,Y_{m})$, i.e.,
the second stage sensors and fusion center have the full information available 
at the first stage aggregator. 
\item[(ii)] {\bf Restricted feedback daisy chain.} Here, we let $U = \gamma_u(Y_1,\ldots,Y_{m})\in \{0,1\}$.
This architecture can be viewed as a parallel configuration in which the fusion center makes a preliminary decision based on the messages from the first $m$ sensors, broadcasts the preliminary decision, and forgets (e.g., due to memory or security constraints) the messages sent by the first $m$ sensors. 
\end{itemize}
\end{itemize}

\subsection{Assumptions and Preliminaries}\label{subsect:Preliminaries}

In this section, we list the basic assumptions that we will be making throughout this paper, and 
note a useful consequence that will be used in our subsequent proofs.

Let $\P_i^X$ be the distribution of a random variable $X$ under hypothesis $H_i$.
Consider the Radon-Nikodym derivative $\ud\P_i^X / \ud\P_j^X$ of the measure
$\P_i^X$ with respect to the measure $\P_j^X$. Informally, this
is the likelihood ratio associated with an observation of $X$, and 
is a random variable whose value is determined by $X$;
accordingly, its value should be denoted by a notation such as
$\Lh{ij}^X(X)$,
where $\Lh{ij}^X$ is a function from $\mathcal{X}$ into $[0,\infty)$ determined by the distributions of $X$ under the two hypotheses. However, in order to avoid cluttered expressions, we
will abuse notation and just write $\Lh{ij}(X)$.
Furthermore, to simplify notation, we use $\Lh{ij}(X,Y)$ in place of $\Lh{ij}((X,Y))$, and similarly for random vectors of arbitrary length. 
We also use $\Lh{ij}(\gamma(X))$ to denote the Radon-Nikodym derivative of the random variable $Z = \gamma(X)$. 
Throughout the paper, we deal with various conditional distributions. 
Abusing notation as before, we let $\Lh{ij}(X | Y)$ be the Radon-Nikodym derivative of the conditional distribution of $X$ given $Y$. Other notations like $\Lh{ij}(\gamma(X) | Y)$ will also be used. 

We make the following assumptions. The first assumption results in no loss of generality (see \cite{TayTsiWin:J07a}). The second assumption is made to simplify the exposition, and can often be relaxed. See \cite{Tsi:88} for a discussion. 

\begin{Assumption}\label{assumpt:EquivalentMeasures}
The measures $\P_0$ and $\P_1$ are absolutely continuous w.r.t.\ each other. 
\end{Assumption}

\begin{Assumption}\label{assumpt:BoundedDivergence}
We have $ \EE{i}{\LLp{ji}{2}(X_1)}< \infty$ for $i,j=0,1$.\footnote{For the Neyman-Pearson formulation, we will only require that this assumption and Lemma \ref{lemma:BoundedDivergence} hold for $i=0,j=1$.}
\end{Assumption}

Assumption \ref{assumpt:BoundedDivergence} implies the following lemma, the proof of which 
follows from Proposition \ref{proposition:JensenDiv} in Appendix \ref{appendix:Mathematical} of \cite{Tsi:88}.

\begin{Lemma}\label{lemma:BoundedDivergence}
There exists some finite constant $a$, such that for all $\gamma\in \Gamma$, and $i,j=0,1$,
\begin{align*}
& \EE{i}{\LLp{ji}{2}(\gamma(X_1))} \leq \EE{i}{\LLp{ji}{2}(X_1)} + 1 < a, \\
& \EE{i}{\left| \LLh{ji}(\gamma(X_1)) \right|} < a.
\end{align*}

\end{Lemma}

\section{Two-Message Architectures}\label{sect:TwoMessage}

In this section, we study the Neyman-Pearson and Bayesian formulations of the decentralized detection problem in two-message architectures. For $i,j \in \{0,1\}$, we consider the log-likelihood ratio at the fusion center, a random variable denoted by $\LLn{ij}$. We have
\begin{align*}
\LLn{ij} &= \LLh{ij}(Y_1^n,Z_1^n)\\
& = \sum_{k=1}^n \LLh{ij}(Y_k) + \LLh{ij}(Z_1^n \mid Y_1^n ) \\
& = \sum_{k=1}^n \LLh{ij}(Y_k) + \sum_{k=1}^n\LLh{ij}(Z_k \mid Y_1^n) \\
& = \sum_{k=1}^n \LLh{ij}(Y_k) + \sum_{k=1}^n \LLh{ij}(Z_k \mid Y_k, W_k).
\end{align*}
The second equality above holds because, under either hypothesis, and given $Y_1^n$, the random variables $Z_k$ are functions of the respective $X_k$; thus, the $Z_k$ are conditionally independent, given $Y_1^k$. The last equality holds because $Z_k$ depends on $Y_1^n$ through $Y_k$ and $W_k$.

To simplify notation, we define, for every possible value $w$ of $W_k$, a random variable  $\LLw{ij}{k}(w)$, according to
\begin{align*}
\LLw{ij}{k}(w) & = \LLh{ij}(Y_k) + \LLh{ij}(Z_k \mid Y_k, W_k = w) \\
& = \LLh{ij}(\gamma_k(X_k), \delta_k^w(X_k)).
\end{align*}
Note that $\LLw{ij}{k}(w)$ is a random variable which is a function of a non-random argument $w$ and the random variable $X_k$. Note also that
$$\LLn{ij} =\sum_{k=1}^n \LLw{ij}{k}(W_k).$$

\subsection{Neyman-Pearson Formulation}\label{subsect:Neyman-Pearson}

Let $\alpha \in (0,1)$ be a given constant. A strategy is {\it admissible} if its Type I error probability satisfies $\P_0(Y_f = 1) < \alpha$. Let $\beta_n^* = \inf \P_1(Y_f=0)$, where the infimum is taken over all admissible strategies for the $n$-sensor problem. Our objective is to characterize the optimal error exponent $\limsup_{n\to\infty}(1/n) \log \beta_n^*$, under different feedback architectures. 

Let $g_{2p}^*$ be the optimal error exponent for the two-message parallel configuration, in which there is no 
feedback from the fusion center, i.e., when each sensor $k$ sends 
two messages, ($\gamma_k(X_k)$, $\delta_k(X_k))$, 
to the fusion center. From \cite{Tsi:88}, the optimal error exponent is
\begin{align*}
g_{2p}^* = \inf_{(\gamma,\delta)\in\Gamma^2} \EE{0}{\LLh{10}(\gamma(X_1), \delta(X_1))}.
\end{align*}
Let $g_{sf}^*$, $g_f^*$, and $g_{rf}^*$ be the optimal error exponents 
for the sequential, full, and restricted feedback architectures respectively. 
Since the sensors can ignore some or all of the feedback messages from the fusion center, we have 
\begin{align}
& g_f^* \leq g_{sf}^* \leq g_{2p}^*, \label{ErrorExpUpperBnd1}\\
& g_f^* \leq g_{rf}^* \leq g_{2p}^*, \label{ErrorExpUpperBnd2}
\end{align}
(Note that error exponents are nonpositive and that smaller error exponents correspond to better performance.)

We will show that under appropriate but mild assumptions, the inequalities in \eqref{ErrorExpUpperBnd1} and \eqref{ErrorExpUpperBnd2} are equalities. Hence, from an asymptotic viewpoint, feedback results in no gain in detection performance. We first show a useful result that underlies a key step in our proofs.

\begin{Lemma}\label{lemma:TypeIILowerBound}
Consider a sequence of strategies, indexed by the number $n$ of sensors, and let 
$\beta_n$ be the associated Type II error probabilities. Let $R$ be a nonnegative constant. 
If
\begin{align*}
\limsup_{n\to\infty} \P_0(\LLn{10} < -nR) < 1 - \alpha,
\end{align*}
then  
\begin{align*}
\liminf_{n\to\infty} \ofrac{n} \log \beta_n \geq -R.
\end{align*}

\end{Lemma}
\begin{IEEEproof}
We have
\begin{align*}
\beta_n & = \P_1(Y_f = 0) \\
& = \E_0\left[ \exp(\LLn{10}) \indicator{Y_f = 0} \right] \\
& \geq \E_0\left[ \exp(\LLn{10}) \indicator{Y_f = 0, \LLn{10} \geq -nR} \right] \\
& \geq e^{-nR} \P_0(Y_f = 0, \LLn{10} \geq -nR).
\end{align*}
Therefore,
\begin{align*}
\P_0(Y_f = 0, \LLn{10} \geq -nR) \leq \beta_n e^{nR}.
\end{align*}
This upper bound yields
\begin{align*}
& 1 - \alpha \leq \P_0(Y_f = 0) \leq \beta_n e^{nR} + \P_0(\LLn{10} < -nR),
\end{align*}
and
\begin{align*}
& \ofrac{n}\log \beta_n + R \geq \ofrac{n}\log(1-\alpha - \P_0(\LLn{10} < -nR)).
\end{align*}
The desired result follows by taking the limit as $n\to\infty$.
\end{IEEEproof}

\subsection{Neyman-Pearson Formulation --- Sequential Feedback}

For the case of sequential feedback, the proof that feedback yields no performance improvement is relatively simple. The core of the proof is an  inequality on the (conditional) expectation of the log-likelihood ratio at the fusion center. We use this inequality together with a variance bound to obtain a bound on the tail probabilities associated with the log-likelihood ratio, and finally use Lemma \ref{lemma:TypeIILowerBound}.

\begin{Theorem}\label{theorem:NP-sequentialFeedback}
Suppose that Assumptions \ref{assumpt:EquivalentMeasures}-\ref{assumpt:BoundedDivergence} hold.
Then, the optimal error exponent for the sequential feedback architecture is $g_{sf}^* = g_{2p}^*$. 
Moreover, there is no loss in optimality if the sensors
ignore the feedback messages from the fusion center and are constrained to 
using the same quantization function. 
\end{Theorem}
\begin{IEEEproof}
From \eqref{ErrorExpUpperBnd1}, we have $g_{sf}^* \leq g_{2p}^*$. 
To show the reverse inequality, we first bound $\E_0[\LLw{10}{k}(W_k) \mid W_k]$ from below by $g_{2p}^*$. We have, for any $w$, 
\begin{align}
\EE{0}{\LLw{10}{k}(W_k) \mid W_k = w} & = 
\EE{0}{\LLh{10}(\gamma_k(X_k), \delta_k^{w}(X_k)) \mid W_k = w} \nonumber\\
& \geq \inf_{(\gamma,\delta)\in\Gamma^2} \EE{0}{\LLh{10}(\gamma(X_1), \delta(X_1))} \nonumber\\
& = g_{2p}^*. \label{iloglhbound}
\end{align}
In particular, $ \EE{0}{\LLw{10}{k}(W_k)} \geq g_{2p}^*$ and
$\EE{0}{\LLn{ij}} \geq n g_{2p}^*$.

We next obtain a suitable variance bound. Let $Q_k = \LLw{10}{k}(W_k) - \E_0[\LLw{10}{k}(W_k)\mid W_k]$. 
From Lemma \ref{lemma:BoundedDivergence}, there exists some constant $a > 0$ such that
\begin{align}
\var_0(Q_k) & \leq \EE{0}{\EE{0}{\left(\LLw{10}{k}(W_k)\right)^2 \mid W_k}} \nonumber\\
& \leq a.\label{ineq:varbound}
\end{align}
Recall that $W_k = Y_1^{k-1}$. We have, for $m < k$, 
\begin{align}
& \EE{0}{Q_m \cdot Q_k} \nonumber\\
& = \EE{0}{Q_m \EE{0}{Q_k \mid W_k}} \nonumber\\
& = 0.\label{ineq:corrbound}
\end{align}

Let $\epsilon > 0$. 
Inequality \eqref{iloglhbound}, together with the bounds 
 \eqref{ineq:varbound} and \eqref{ineq:corrbound}, and Chebyshev's inequality, yield
\begin{align*}
\P_0\left( \LLn{10} < n(1+\epsilon)g_{2p}^* \right) 
& \leq \P_0\left( \sum_{k=1}^n Q_k < n\epsilon g_{2p}^* \right) \\
& \leq \frac{a}{n\epsilon^2 (g_{2p}^*)^2}.
\end{align*}
Letting $n\to\infty$, we get
$$
\lim_{n\to\infty}\P_0\left( \LLn{10} < n(1+\epsilon)g_{2p}^* \right) = 0 < 1-\alpha.
$$
Therefore, applying Lemma \ref{lemma:TypeIILowerBound}, we have $g_{sf}^* \geq (1+\epsilon)g_{2p}^*$. 
Since $\epsilon$ was chosen arbitrarily, we obtain $g_{sf}^* \geq g_{2p}^*$, and the proof is complete.
\end{IEEEproof}

\subsection{Neyman-Pearson Formulation --- Full Feedback}
Next, we consider the full feedback architecture. The same architecture has been studied in \cite{ShaPap:93}, using the method of types, and under a more restrictive set of assumptions. In the following, we show a result similar to the one in \cite{ShaPap:93}, i.e., that there is no gain from the feedback messages asymptotically. For a comparison, we note that  \cite{ShaPap:93} involved a constraint that feedback messages take values in an alphabet that grows at most subexponentially. This constraint excludes the full feedback case, in which the feeback messages $W_k$ take values in an exponentially growing alphabet.

The following result subsumes, in some sense Theorem \ref{theorem:NP-sequentialFeedback}; indeed, if full feedback cannot improve performance, then sequential feedback cannot either. On the other hand, for this more general result we will need a stronger assumption. In Theorem 
\ref{theorem:NP-sequentialFeedback}, we used the property that the ``innovations'' $\LLw{10}{m}(W_m) - \E_0[\LLw{10}{m}(W_m)\mid W_m]$ were uncorrelated, which allowed us to use Chebyshev's inequality. Such a property is no longer true in the full feedback case. Instead, we 
impose an exponential tail bound on the original log-likelihood ratios;  equivalently, we make a finiteness assumption on the log moment generating function of the original log-likelihood ratios about a neighborhood of the origin, which is standard in the theory of large deviations \cite{DemZei:98}. We then proceed to derive related bounds that refer to the log-likelihood ratios associated with various messages. This step is somewhat tedious but unsurprising.

We define $b(s)=\log \E_0[(\Lh{10}(X_1))^s]$, which is the log moment generating function of the log-likelihood ratio $\log \Lh{10}(X_1))$.

\begin{Assumption}\label{assumpt:BoundedLH}
There exists some $\bar{s} < 0$ such that $b(\bar{s})<\infty$.
\end{Assumption}

Since $b(\cdot)$ is nonincreasing on $[\bar{s},0]$ (cf.\ Lemma 2.2.5 of \cite{DemZei:98}), Assumption \ref{assumpt:BoundedLH} implies that $b(s) < \infty$ for all $s \in [\bar{s},0]$. Furthermore, the second moment of $\LLh{10}(X_1)$ under $H_0$ exists and is finite. Therefore, Assumption \ref{assumpt:BoundedLH} implies Assumption \ref{assumpt:BoundedDivergence} for $i=0$ and $j=1$.
 
Consider a pair $\xi = (\gamma, \delta) \in \Gamma^2$ of quantization functions. 
Let $\varphi_{\xi}(s) = \log \EE{0}{(\Lh{10}(\xi(X_1))^s}$ be the log moment generating function of the log-likelihood ratio of the distribution of $\xi(X_1) = (\gamma(X_1), \delta(X_1))$ under $H_1$ versus that under $H_0$. 
Suppose that a strategy sequence has been fixed. Let
$\psi_n(s) =  \log \E_0[\exp(s \LLn{10})]$ be the log moment generating function of $\LLn{10}$.

Based on Assumption \ref{assumpt:BoundedLH}, we will show some properties of $\varphi_\xi$ and $\psi_n$. We will then use these properties to obtain tail bounds on  $\LLn{10}$, which will play the same role as the Chebyshev bound in the proof of Theorem \ref{theorem:NP-sequentialFeedback}.
\footnote{Throughout the paper, we use $f'(s)$ and $f''(s)$ to denote the first and second derivatives of $f$ w.r.t.\ $s$.}

\begin{Lemma}\label{lemma:BoundedLogMoment}
Suppose Assumption \ref{assumpt:EquivalentMeasures}-\ref{assumpt:BoundedLH} holds, and let $\bar{s}$ be as in Assumption \ref{assumpt:BoundedLH}. 
\begin{enumerate}[(i)]
\item\label{it:varphi2deriv} There exists a positive constant $c$ such that for all $s \in [\bar{s}/2,0]$, and for all $\xi\in\Gamma^2$, we have $0 \leq \varphi_\xi''(s) \leq c$.
\item\label{it:varphibound} Let $\xi^* \in \Gamma^2$ be such that $\varphi'_{\xi^*}(0) \leq g_{2p}^* + \epsilon$, where $\epsilon$ is a small positive constant so that $h = \sqrt{\epsilon/(2c)} < \min\{|\bar{s}|/2, 1/4\}$. Then, for all $s\in [-h,0]$, and for all $\xi\in\Gamma^2$, we have $\varphi_\xi(s) \leq \varphi_{\xi^*}(s) + \epsilon/2$. 
\item\label{it:NPpsi} For all $n \geq 1$ and $s\in [-h,0]$, we have $\psi_n(s)/n\leq \varphi_{\xi^*}(s) + \epsilon$. 
\end{enumerate}
\end{Lemma}
\begin{IEEEproof}
We first prove claim \eqref{it:varphi2deriv}. From Lemma 2.2.5 of \cite{DemZei:98}, $\varphi_\xi$ is a convex function with nonnegative second derivatives. We next show that its second derivative is uniformly upper bounded for all $\xi\in\Gamma^2$. From Lemma \ref{proposition:JensenDiv}, we have
\begin{align}
\EE{0}{(\Lh{10}(\xi(X_1)))^s}
& \leq \EE{0}{(\Lh{10}(X_1))^s} \leq e^{b(\bar{s})}. \label{ineq:uniformb}
\end{align}
Let $f(s) = \EE{0}{(\Lh{10}(\xi(X_1)))^s}$, and $\eta = \min\{|\bar{s}|/2,1\}$. There exists a positive constant $M$ such that for all $|x| > M$, we have $x^2 \leq \exp(\eta x) + \exp(-\eta x)$. Making use of this bound, we obtain
\begin{align*}
\varphi''_\xi(s) 
& = \frac{\EE{0}{(\Lh{10}(\xi(X_1)))^s \log^2\Lh{10}(\xi(X_1))}}{\EE{0}{(\Lh{10}(\xi(X_1)))^s}} - (\varphi'_\xi(s))^2 \\
& \leq \EE{0}{(\Lh{10}(\xi(X_1)))^s \log^2\Lh{10}(\xi(X_1))} \\
& \leq M^2 f(s) + f(s+\eta) + f(s-\eta) \\
& \leq (M^2+2)f(\bar{s}) \\
& \leq (M^2+2)e^{b(\bar{s})}.
\end{align*}
The third inequality follows from the bounds $\bar{s} < s+\eta < 1$ and $\bar{s} < s - \eta <0 $, and the facts that $f(x)$ is nonincreasing over $[\bar{s},0]$, while $f(x) \leq 1 \leq f(\bar{s})$ for $x \in [0,1]$. The final inequality follows from \eqref{ineq:uniformb}. Claim \eqref{it:varphi2deriv} is now proved. 

We now use a Taylor series expansion to prove claim \eqref{it:varphibound}. Since $\varphi_\xi(0) = 0$ for any $\xi \in \Gamma^2$, we have for $s\in [-h,0]$, 
\begin{align*}
\varphi_\xi(s) - \varphi_{\xi^*}(s) 
& = (\varphi'_\xi(0) - \varphi'_{\xi^*}(0)) s + (\varphi''_\xi(s_1) - \varphi''_{\xi^*}(s_2)) \frac{s^2}{2} \\
& \leq \epsilon |s| + c \frac{s^2}{2} \\
& \leq \epsilon/2,
\end{align*}
where $s_1$ and $s_2$ are between $s$ and $0$, and the first inequality follows from $\varphi'_\xi(0) \geq g_{2p}^*$, $\varphi'_{\xi^*}(0) \leq g_{2p}^* + \epsilon$, $\varphi''_\xi(s_1) \leq c$, and $\varphi''_{\xi^*}(s_2) \geq 0$.

Finally, we turn to the proof of claim \eqref{it:NPpsi}. Recall that $Z_k = \delta_k^{W_k}(X_k)$.
For $s\in [-h,0]$, we have
\begin{align}
& \EE{0}{\prod_{k=1}^n (\Lh{10}(Z_k | Y_k, W_k))^{s} \Bigmid Y_1^n} \nonumber\\
& = \EE{0}{\prod_{k=1}^n (\Lh{10}(\delta_k^{W_k}(X_k) | Y_k))^{s} \Bigmid Y_1^n} \nonumber\\
& = \prod_{k=1}^n \EE{0}{(\Lh{10}(\delta_k^{W_k}(X_k) | Y_k))^{s} \Bigmid Y_1^n} \nonumber\\
& \leq \prod_{k=1}^n \left(\EE{0}{(\Lh{10}(\delta^{Y_k}(X_k) | Y_k))^{s} \Bigmid Y_k} + \epsilon_1 \right),\label{ineq:innerbnd}
\end{align}
where $\epsilon_1 = \epsilon \exp(-b(\bar{s}))/2$, and $\delta^{Y_k} \in \Gamma$ is a function depending on the value of $Y_k$, and is such that 
\begin{align*}
\EE{0}{(\Lh{10}(\delta^{Y_k}(X_k) | Y_k))^{s} \Bigmid Y_k}
\geq \sup_{\delta\in\Gamma} \EE{0}{(\Lh{10}(\delta(X_k) | Y_k))^{s} \Bigmid Y_k} - \epsilon_1.
\end{align*}
From \eqref{ineq:innerbnd}, we have
\begin{align*}
\frac{\psi_{n}(s)}{n} & = \ofrac{n}\log \EE{0}{(\Lh{10}(Y_1^n))^{s} \cdot
\EE{0}{\prod_{k=1}^n (\Lh{10}(Z_k | Y_k, W_k))^{s} \Bigmid Y_1^n}} \\
& \leq \ofrac{n}\log \EE{0}{(\Lh{10}(Y_1^n))^{s} \cdot
\prod_{k=1}^n \left(\EE{0}{(\Lh{10}(\delta^{Y_k}(X_k) | Y_k))^{s} \Bigmid Y_k} + \epsilon_1 \right)} \\
& = \ofrac{n}\log \prod_{k=1}^n \Big(\EE{0}{(\Lh{10}(Y_k,\delta^{Y_k}(X_k)))^{s}} + \epsilon_1 \EE{0}{(\Lh{10}(Y_k))^s} \Big) \\
& \leq \ofrac{n}\sum_{k=1}^n\log \EE{0}{(\Lh{10}(\gamma_k(X_k),\delta^{Y_k}(X_k)))^s} + \frac{\epsilon}{2},
\end{align*}
where the last inequality follows from \eqref{ineq:uniformb}, and the inequality $\log(x+\epsilon) \leq \log x + \epsilon$ for $x \geq 1$. 
Let $\xi_k \in \Gamma^2$ such that $\xi_k(X_k) = (\gamma_k(X_k), \delta_k(X_k))$, where $\delta_k(X_k) = \delta^{u}(X_k)$ iff $\gamma_k(X_k) = u \in \mathcal{T}$. 
We therefore have
\begin{align*}
\frac{\psi_{n}(s)}{n} 
& \leq \ofrac{n}\sum_{k=1}^n \varphi_{\xi_k}(s) + \frac{\epsilon}{2} \leq \varphi_{\xi^*}(s) + \epsilon,
\end{align*}
where the second inequality follows from claim \eqref{it:varphibound}. The proof is now complete.
\end{IEEEproof}

Finally, we show that for both the full and restricted feedback architectures, feedback does not improve the optimal error exponent.
\begin{Theorem}\label{theorem:NP-FullFeedback}
Suppose that Assumptions \ref{assumpt:EquivalentMeasures}-\ref{assumpt:BoundedLH} hold.
Then, in both the full and restricted feedback architectures, there is no loss in optimality if sensors
ignore the feedback messages from the fusion center, i.e., $g_{f}^* = g_{rf}^* = g_{2p}^*$.
Moreover, there is no loss in optimality if all sensors are constrained to 
using the same quantization function. 
\end{Theorem}
\begin{IEEEproof}
From \eqref{ErrorExpUpperBnd2}, it suffices to show $g_f^* \geq g_{2p}^*$.
Choose a sufficiently small $\epsilon > 0$. Let $\xi^*$ and $h$ be chosen as in Lemma \ref{lemma:BoundedLogMoment}\eqref{it:varphibound}, and let $t_{\epsilon} = -(\varphi_{\xi^*}(-h) + \epsilon)/h$. From the Chernoff bound and Lemma \ref{lemma:BoundedLogMoment}\eqref{it:NPpsi}, we have 
\begin{align*}
\limsup_{n\to\infty} \ofrac{n} \log \P_0\left( \LLn{10} < n(t_{\epsilon}-\epsilon) \right) 
& \leq \varphi_{\xi^*}(-h) + \epsilon + h(t_{\epsilon}- \epsilon)\\
& = -h \epsilon < 0.
\end{align*}
Applying Lemma \ref{lemma:TypeIILowerBound}, we have $g_f^* \geq t_{\epsilon}-\epsilon$. 
The Taylor series expansion of $\varphi_{\xi^*}$ yields
\begin{align*} 
t_{\epsilon} = \varphi'_{\xi^*}(0) - \varphi''_{\xi^*}(\theta)\ofrac{2}\sqrt{\frac{\epsilon}{2c}} - \sqrt{2c\epsilon},
\end{align*}
where $\theta \in [-h, 0]$, and $c$ is the same constant as in Lemma \ref{lemma:BoundedLogMoment}\eqref{it:varphi2deriv}. Since $0 \leq \varphi''_{\xi^*}(\theta) \leq c$, $t_{\epsilon} \to g_{2p}^*$ as $\epsilon$ decreases to 0. Letting $\epsilon\to 0$, we obtain the theorem. 
\end{IEEEproof}

\subsection{Bayesian Formulation}\label{subsect:Bayesian}

In this section, we show that feedback does not improve the optimal error exponent for the binary Bayesian decentralized detection problem in the sequential, full, and restricted feedback architectures. Let the prior probability of hypothesis $H_j$ be $\pi_j > 0$, $j=0,1$. Given a strategy, the probability of error at the fusion center is $P_e(n) = \pi_0 \P_0(Y_f = 1) + \pi_1 \P_1(Y_f = 0)$.
Let $P_e^*(n)$ be the minimum  probability of error, over all strategies, for the $n$-sensor problem.
We seek to characterize the optimal error exponent
\begin{align*}
\limsup_{n \to \infty} \ofrac{n} \log P_e^*(n).
\end{align*}
From \cite{Tsi:88}, the optimal error exponent for the parallel configuration 
without any feedback is given by 
\begin{align}\label{eq:optimalParallel_Bayesian}
\mathcal{E}_{2p}^* = \inf_{(\gamma,\delta)\in\Gamma^2} \min_{s \in [0,1]} 
\log \EE{0}{\left(\Lh{10}(\gamma(X_1),\delta(X_1))\right)^s}.
\end{align}
Similar to the Neyman-Pearson formulation, we let $\mathcal{E}_{sf}^*$, $\mathcal{E}_{rf}^*$ and $\mathcal{E}_{f}^*$ denote the optimal error exponents for the sequential, restricted, and full feedback architectures respectively. Note that the counterparts of inequalities \eqref{ErrorExpUpperBnd1} and \eqref{ErrorExpUpperBnd2} also hold for the Bayesian error exponents. Therefore, to show that feedback does not improve the asymptotic performance, it suffices to show a lower bound for the full feedback architecture. Recall that $\psi_{n}(s) = \log \EE{0}{\exp(s\LLn{10})}$ is the log moment generating function of $\LLn{10}$. 
The following lemma provides uniform bounds for $\psi_n$ and its derivatives, over all strategies.

\begin{Lemma}\label{lemma:psi}
Suppose that Assumptions \ref{assumpt:EquivalentMeasures} and \ref{assumpt:BoundedDivergence} hold.
\begin{enumerate}[(i)]
\item\label{it:bnd1der} For all $s\in [0,1]$, we have $\EE{0}{\log\Lh{10}(X_1)} \leq \psi_{n}'(s)/n \leq \EE{1}{\log\Lh{10}(X_1)}$. 
\item\label{it:bnd2der} For any bounded sequence $(t_n)$ and for any given strategy such that there exists $s_n\in (0,1)$ with $\psi_n'(s_n) = t_n$ for each $n$,\footnote{Note that the sequence $(s_n)$ depends on the strategy used.}, we have $\psi_{n}''(s_n) \leq nC$, where $C$ is a constant independent of the strategy.
\item\label{it:psilb} For all $s \in [0,1]$, we have $\psi_{n}(s) \geq n\mathcal{E}_{2p}^*$.
\end{enumerate}
\end{Lemma}
\begin{IEEEproof}
We first show claim \eqref{it:bnd1der}.
To show the bounds on $\psi_n'(s)$, we note that $\psi_n$ is convex, so $\psi_n'(0) \leq \psi_n'(s) \leq \psi_n'(1)$ for all $s \in [0,1]$. Using Proposition \ref{proposition:JensenDiv}, it is then easy to check that $\psi_n'(0) \geq n\EE{0}{\log\Lh{10}(X_1)}$ and $\psi_n'(1) \leq n\EE{1}{\log\Lh{10}(X_1)}$.
\

Next, we prove claim \eqref{it:bnd2der}. We have
\begin{align}
\psi_n''(s_n)
& = \frac{\E_0[(\LLn{10})^2\exp(s_n\LLn{10})]}{\E_0[\exp(s_n\LLn{10})]} - (\psi_n'(s_n))^2 \nonumber\\
& \leq C_1 \E_0[(\LLn{10})^2\exp(s_n\LLn{10})], \label{ineq:psibound1}
\end{align}
where the inequality follows from the bound $\E_0[\exp(s_n\LLn{10})] \geq 1/C_1$, for some constant $C_1$ independent of the strategy. (This fact is proved in Proposition 3 of \cite{Tsi:88}.)
The right-hand side of \eqref{ineq:psibound1} can be upper bounded by observing that
\begin{align}
\E_0[(\LLn{10})^2\exp(s_n\LLn{10})]
& = \EE{0}{(\LLn{10})^2 e^{s_n\LLn{10}}\indicator{\LLn{10}\leq 0}} 
+ \EE{1}{(\LLn{10})^2 e^{-(1-s_n)\LLn{10}}\indicator{\LLn{10}> 0}} \nonumber\\
& \leq 4\left(\ofrac{s_n^2} + \ofrac{(1-s_n)^2}\right),\label{ineq:psibound2}
\end{align}
where in the inequality, we use the result that the function $f_1(x) = x^2\exp(s_n x) \indicator{x\leq 0}$ is maximized at $-2/s_n$, and the function $f_2(x) = x^2\exp(-(1-s_n) x) \indicator{x > 0}$ is maximized at $2/(1-s_n)$. It now suffices to show that both $s_n$ and $1-s_n$ are at least $C_2/\sqrt{n}$ for some constant $C_2$ independent of the particular strategy chosen. To simplify the notation, let $\ell_n = \exp(\LLn{10})$. Suppose that $|t_n| \leq t$ for all $n$.
Using the inequalities $x^s \leq sx + 1$ for $0< s < 1$, and $x^s \geq x$ for $x \leq 1$, we obtain from the equation $\psi'_n(s_n) = t_n$,\footnote{We use the notations $x^+ = \max(x,0)$ and $x^- = - \min(x,0).$}
\begin{align*}
t_n \EE{0}{\exp(s_n \LLn{10})} & = \EE{0}{(\ell_n)^{s_n} \log \ell_n} \\
& = \EE{0}{(\ell_n)^{s_n} (\log \ell_n)^+} - \EE{0}{(\ell_n)^{s_n} (\log \ell_n)^-} \\
& \leq s_n \left(\EE{0}{\ell_n(\log \ell_n)^+} + \EE{0}{(\log \ell_n)^+} \right) 
- \EE{0}{\ell_n(\log \ell_n)^-},
\end{align*}
which yields
\begin{align}\label{ineq:snbound}
s_n & \geq \frac{\EE{1}{(\log \ell_n)^-} - t}{\EE{1}{(\log \ell_n)^+} + \EE{0}{(\log \ell_n)^+}},
\end{align}
since $0 \leq \EE{0}{\exp(s_n \LLn{10})} \leq 1$ and $|t_n| \leq t$. 
We first bound the denominator in \eqref{ineq:snbound} by using
$g(x) = x(\log x)^+$, which is a convex function, and Proposition \ref{proposition:JensenDiv} to get
\begin{align}
\EE{1}{(\log \ell_n)^+} 
& = \EE{0}{g(\ell_n)} \nonumber\\
& \leq \EE{0}{g(\Lh{10}(X_1^n))} \nonumber\\
& \leq \EE{0}{\Lh{10}(X_1^n) |\log \Lh{10}(X_1^n)|} \nonumber\\
& = \EE{1}{|\log \Lh{10}(X_1^n)|} \nonumber\\
& \leq \sum_{k=1}^n \EE{1}{|\log \Lh{10}(X_k)|} \nonumber\\
& \leq \sum_{k=1}^n \EE{1}{\log^2 \Lh{10}(X_k)} + n \nonumber\\
& \leq n C_3, \label{ineq:dembound}
\end{align}
where $C_3$ is a constant, and the last inequality follows from Lemma \ref{lemma:BoundedDivergence}.
Similarly, it can be shown that $\EE{0}{(\log \ell_n)^+} = \E_0[(\LLn{01})^-] \leq \E_0[(\LLn{01})^+]$ is bounded by $nC_3$. 
Next, we show a lower bound for the numerator in \eqref{ineq:snbound}.
Let 
\begin{align*}
f(x) 
= \left \{ \begin{array}{ll}
x(\log x)^-, & {\rm if}\ 0 \leq x \leq 1, \\
1 - x, & {\rm if}\ x > 1,
\end{array} \right.
\end{align*}
which is a concave function not greater than $x(\log x)^-$. From Proposition \ref{proposition:JensenDiv}, we obtain
\begin{align}
\EE{1}{(\log \ell_n)^-}
& = \EE{0}{\ell_n(\log \ell_n)^-} \nonumber\\
& \geq \EE{0}{f(\ell_n)} \nonumber\\
& \geq \EE{0}{f(\Lh{10}(X_1^n))} \nonumber\\
& \geq \EE{1}{(\log \Lh{10}(X_1^n))^-} - \P_1(\Lh{10}(X_1^n) > 1) \nonumber\\
& \geq \sqrt{n} \EE{1}{\left( \ofrac{\sqrt{n}}\sum_{k=1}^n \log \Lh{10}(X_k) \right)^-} - 1. \nonumber
\end{align} 
Applying Fatou's Lemma and the Central Limit Theorem, we obtain
\begin{align}
\liminf_{n\to\infty} \frac{\EE{1}{(\log \ell_n)^-}}{\sqrt{n}}
& \geq C_4, \label{ineq:numbound}
\end{align} 
where $C_4$ is a positive constant.
Substituting the bounds \eqref{ineq:dembound} and \eqref{ineq:numbound} into \eqref{ineq:snbound}, we finally have
\begin{align*}
\liminf_{n\to\infty} s_n \sqrt{n} \geq C_2,
\end{align*}
for some positive constant $C_2$. 
A similar proof using 
$$
\E_1[\LLn{01}\exp((1-s_n)\LLn{01})] = - \E_0[\LLn{10}\exp(s_n\LLn{10})] 
= - t_n \EE{0}{\exp(s_n \LLn{10})} \geq - t
$$ 
shows that the same bound holds for $1-s_n$. 
Therefore, from \eqref{ineq:psibound2}, claim \eqref{it:bnd2der} holds.
\

In the following, we establish claim \eqref{it:psilb}. Let $\epsilon$ be a positive constant.
Similar to the proof of Lemma \ref{lemma:BoundedLogMoment}\eqref{it:NPpsi}, let $\delta^{Y_k} \in \Gamma$ be a function depending on the value of $Y_k$, so that 
\begin{align*}
\EE{0}{(\Lh{10}(\delta^{Y_k}(X_k) | Y_k))^{s} \Bigmid Y_k}
\leq \inf_{\delta\in\Gamma} \EE{0}{(\Lh{10}(\delta(X_k) | Y_k))^{s} \Bigmid Y_k} + \epsilon.
\end{align*}
We have
\begin{align}
\psi_{n}(s) & = \log \EE{0}{(\Lh{10}(Y_1^n))^{s} \cdot
\EE{0}{\prod_{k=1}^n (\Lh{10}(Z_k | Y_k, W_k))^{s} \Bigmid Y_1^n}} \nonumber\\
& \geq \ofrac{n}\log \EE{0}{(\Lh{10}(Y_1^n))^{s} \cdot
\prod_{k=1}^n \left(\EE{0}{(\Lh{10}(\delta^{Y_k}(X_k) | Y_k))^{s} \Bigmid Y_k} - \epsilon \right)} \nonumber\\
& = \sum_{k=1}^n \log \EE{0}{(\Lh{10}(Y_k))^{s}
\left(\EE{0}{(\Lh{10}(\delta^{Y_k}(X_k) | Y_k))^{s} \Bigmid Y_k} - \epsilon \right)
} \nonumber\\
& \geq \sum_{k=1}^n \log \left(\EE{0}{(\Lh{10}(Y_k, \delta^{Y_k}(X_k)))^{s}} - \epsilon \right), \label{ineq:psinbound1}
\end{align}
where we have used the inequality $\EE{0}{(\Lh{10}(Y_k))^{s}} \leq 1$ in \eqref{ineq:psinbound1}.
Recall that $Y_k = \gamma_k(X_k)$. We can define $\xi_k \in \Gamma^2$ such that $\xi_k(X_k) = (\gamma_k(X_k), \delta_k(X_k))$, where $\delta_k(X_k) = \delta^{u}(X_k)$ iff $\gamma_k(X_k) = u \in \mathcal{T}$. 
From \eqref{ineq:psinbound1}, we obtain the bound
\begin{align}
\psi_n(s)
& \geq \sum_{k=1}^n \log \left(
\EE{0}{(\Lh{10}(\xi_k(X_k)))^{s}} - \epsilon \right) \nonumber\\
& \geq n\log \left(\inf_{\xi \in \Gamma^2} \EE{0}{(\Lh{10}(\xi(X_1)))^{s}} - \epsilon \right). \label{ineq:momentbound}
\end{align}
Since $\epsilon$ is arbitrary, the lemma is proved.
\end{IEEEproof}

\begin{Theorem}\label{theorem:Bayesian-FullFeedback}
Suppose that Assumptions \ref{assumpt:EquivalentMeasures} and \ref{assumpt:BoundedDivergence} hold.
Then $\mathcal{E}_{f}^* = \mathcal{E}_{2p}^*$.
Moreover, there is no loss in optimality if sensors are constrained to 
using the same quantization function, which
ignore the feedback messages from the fusion center.
\end{Theorem}
\begin{IEEEproof}
It is clear that $\mathcal{E}_{f}^* \leq \mathcal{E}_{2p}^*$. 
To show the reverse bound, we make use of Proposition \ref{proposition:errorproblowerbound}.
Let the conditional probability of error under $H_j$ be $P_{n,j}$ for $j=0,1$.
Let $s^*_n = \arg\min_{s\in (0,1)} \psi_n(s)$ so that $\psi_n'(s^*_n) = 0$. From Proposition  \ref{proposition:errorproblowerbound}, we have
\begin{align*}
\max_{j=0,1} P_{n,j} 
& \geq \ofrac{4} \exp\left(\psi_n(s^*_n) - \sqrt{2\psi_n''(s^*_n)}\right) \\
& \geq \exp(\psi_n(s^*_n) - C \sqrt{n}) \\
& \geq \exp(n \mathcal{E}_{2p}^* - C \sqrt{n})
\end{align*}
where $C$ is some constant. The penultimate inequality follows from Lemma \ref{lemma:psi}\eqref{it:bnd2der}, and the last inequality from Lemma \ref{lemma:psi}\eqref{it:psilb}. Letting $n\to \infty$, we have
\begin{align*}
\limsup_{n\to\infty} \ofrac{n}\log P_e(n)
& = \limsup_{n\to\infty} \ofrac{n} \log \max_{j=0,1} P_{n,j} \\
& \geq \mathcal{E}_{2p}^*.
\end{align*}
This implies that $\mathcal{E}_{f}^* \geq \mathcal{E}_{2p}^*$, 
and the proof is complete.
\end{IEEEproof}

Since the sequential and restricted feedback configurations can perform no better than the full feedback architecture, and no worse than the parallel configuration, we have the following result.

\begin{Theorem}\label{theorem:Bayesian-SequentialFeedback}
Suppose that Assumptions \ref{assumpt:EquivalentMeasures} and \ref{assumpt:BoundedDivergence} hold.
Then $\mathcal{E}_{sf}^* = \mathcal{E}_{rf}^* = \mathcal{E}_{2p}^*$.
Moreover, there is no loss in optimality if sensors are constrained to 
using the same quantization function, which ignore the feedback messages from the fusion center.
\end{Theorem}

\section{One-Message Architectures}\label{sect:OneMessage}

In this section, we consider the one-message architecture. We study both the Neyman-Pearson and Bayesian formulations for the binary hypothesis testing problem. Similar to the two-message architecture, feedback in general does not improve the asymptotic detection performance, except for the case of Bayesian detection with restricted feedback in the daisy chain architecture. In the case where there is no feedback \cite{Tsi:88}, the optimal Neyman-Pearson error exponent is 
\begin{align*}
g_{1p}^* = \inf_{\gamma\in\Gamma} \EE{0}{\LLh{10}(\gamma(X_1))},
\end{align*}
while the optimal Bayesian error exponent is
\begin{align*}
\mathcal{E}_{1p}^* = \inf_{\gamma\in\Gamma} \min_{s \in [0,1]} 
\log \EE{0}{\left(\Lh{10}(\gamma(X_1))\right)^s}.
\end{align*}

\subsection{Full Information at Fusion Center}\label{OneMessageFull}

We consider the case where the fusion center has access to all sensor messages. This is the case for the sequential feedback architecture in which the fusion center is the last sensor. The same applies for the full feedback daisy chain architecture. By ignoring all feedback messages except at the fusion center, these architectures are equivalent to the parallel configuration with the same number of sensors. Therefore, the optimal error exponents under both the Neyman-Pearson and Bayesian formulations are at least as negative as those for the parallel configuration. The proof of the reverse direction involves the same steps as in the proofs for the two-message architectures in Section \ref{sect:TwoMessage}. Specifically, the proof for the one-message sequential feedback architecture is similar to that of Theorem \ref{theorem:NP-sequentialFeedback}, with suitable modifications (remove all references to the first messages $\gamma_k$ and replace $Y_k$ by $Z_k$). The proof for the daisy-chain architecture corresponds to that of Theorems \ref{theorem:NP-FullFeedback} and \ref{theorem:Bayesian-FullFeedback}. The result for the daisy-chain architecture under the Neyman-Pearson formulation is also provided in \cite{Zou:Msc09}. The above discussion is summarized in the following result, whose proof is omitted. 

\begin{Theorem}\label{theorem:OneMessageFull}
Suppose that Assumptions \ref{assumpt:EquivalentMeasures} and \ref{assumpt:BoundedDivergence} hold.
\begin{enumerate}
\item Under either the Neyman-Pearson or Bayesian formulation, the optimal error exponents for the one-message sequential feedback are the same as that of the parallel configuration under either corresponding formulations. 
\item Under the Bayesian formulation, the optimal error exponent for the full feedback daisy chain is the same as that of the parallel configuration. In addition, if Assumption \ref{assumpt:BoundedLH} holds, the Neyman-Pearson error exponent is the same as that of the parallel configuration.
\item Furthermore, there is no loss in optimality if all sensors (except the last sensor in the one-message sequential feedback architecture) are constrained to using the same quantization function. 
\end{enumerate}
\end{Theorem}

\subsection{Restricted Feedback Daisy Chain}\label{DaisyChainRestricted}

In this section, we consider the restricted feedback daisy chain (RFDC) architecture. 
References \cite{Zou:Msc09,KreTsiZou:09} have shown that under the Neyman-Pearson formulation, feedback again does not improve the optimal error exponent. In this section, we consider the Bayesian formulation, and show that unlike the Neyman-Pearson formulation, feedback may improve the detection performance. We provide a characterization of the optimal error exponent in this case.

We assume that $\lim_{n\to\infty} m/n = r \in (0,1)$, otherwise the architecture
is equivalent to a parallel configuration. Let $\mathcal{E}^*_{dc}$ be the optimal error exponent.
For $\gamma\in\Gamma$, and $j=0,1$, let the Fenchel-Legendre transform of the log moment generating functions be
\begin{align*}
\Lambdas{j}{\gamma, t} & = \sup_{s\in\Real} \left\{ s t - \log \EE{j}{e^{s\LLh{10}(\gamma(X_1))}} \right\}.
\end{align*}
These are also known as rate functions \cite{DemZei:98}. 
For $i,j \in \{0,1\}$, any for any given sequence of strategies for the first $m$ sensors, let the rate of decay of the conditional probabilities be
\begin{align*}
e_{ij} & = -\limsup_{n\to\infty} \ofrac{m}\log \P_i(U=j).
\end{align*}
We collect the decay rates into a vector 
\begin{align}\label{eq:vece}
\vec{e} = [e_{01},e_{10},e_{00},e_{11}].
\end{align}

\begin{Lemma}\label{lemma:RestrictedFBDCLowerBound}
Suppose Assumptions \ref{assumpt:EquivalentMeasures} and \ref{assumpt:BoundedDivergence} hold.
Suppose that the quantization functions for sensors $1,\ldots,m$ in a RFDC have been fixed. 
Then, we have
\begin{align*}
\limsup_{n\to\infty} \ofrac{n} \log P_e(n) 
& = - h(\vec{e})
\end{align*}
where $\vec{e}$ is as defined in \eqref{eq:vece},
\begin{align}
h(\vec{e}) = \min\Big\{
& (1-r)\sup_{\delta^0\in\Gamma} \Lambdas{0}{\delta^0, \frac{r}{1-r}(e_{10} - e_{00})} + re_{00}, \nonumber\\
& (1-r)\sup_{\delta^1\in\Gamma} \Lambdas{1}{\delta^1, -\frac{r}{1-r}(e_{01} - e_{11})} + re_{11} 
\Big\},\label{eq:h}
\end{align}
and $P_e(n)$ is the optimal probability of error under the given quantization functions for sensors $1,\ldots,m$.
\end{Lemma}
\begin{IEEEproof}
Let us fix a sequence of strategies that conform to the given 
quantization functions for sensors $1,\ldots,m$.
Let $\hat{\alpha}_n$ and $\hat{\beta}_n$ be the Type I and II error probabilities 
of a strategy with the fusion rule 
\begin{align*}
Y_f
= \left \{ 
\begin{array}{ll}
0, & {\rm if}\ \LLn{10} \leq 0, \\
1, & {\rm if}\ \LLn{10} > 0.
\end{array} 
\right.
\end{align*}

From the Neyman-Pearson Lemma \cite{LehRom:B05}, the optimal decision rule at the fusion center is the Neyman-Pearson test. Moreover, for any given fusion rule, either the Type I or II error probability is at least $\hat{\alpha}_n$ or $\hat{\beta}_n$. Therefore, we have
\begin{align}
\limsup_{n\to\infty} \ofrac{n}\log P_e(n)
& \geq \min\{ \limsup_{n\to\infty} \ofrac{n} \log \hat{\alpha}_n, 
\limsup_{n\to\infty} \ofrac{n} \log \hat{\beta}_n \}. \label{eq:PelowerBnd}
\end{align}
Thus it suffices to find a lower bound for the strategy using a zero threshold
log likelihood ratio test as a fusion rule. Henceforth, we will assume that such a fusion rule is employed. 
Conditioning on the value of $U$, we have
\begin{align*}
P_1(Y_f = 0) = \P_1(Y_f=0\mid U = 0)\P_1(U=0) + \P_1(Y_f=0\mid U =1)\P_1(U=1).
\end{align*}

Fix an $\epsilon > 0$. Let $\delta_i(\cdot, u) = \delta_i^u(\cdot) \in \Gamma$ be a function that depends on the value of $u$. Let $l = n - m$. Using the lower bound in Cram\`er's Theorem (cf.\ \cite{Str:93}) and Lemma \ref{lemma:psi}, we obtain 
\begin{align*}
& \ofrac{n} \log \P_1(Y_f=0\mid U = 0) \\
& = \ofrac{n} \log \P_1(\LLn{10}/l \leq 0 \mid U = 0) \\
& = \ofrac{n} \log \P_1\left(\ofrac{l} \sum_{k=1}^{l} \LLh{10}(\delta_k^0(X_k)) 
\leq -\ofrac{l} \log \frac{\P_1(U=0)}{\P_0(U=0)}\right) \\
& \geq - \frac{l}{n}\cdot\ofrac{l} \sum_{k=1}^{l} \Lambdas{1}{\delta_k^0, 
-\ofrac{l} \log \frac{\P_1(U=0)}{\P_0(U=0)}-\epsilon}+ o(1) \\
& \geq - \frac{l}{n} \sup_{\delta^0\in\Gamma} \Lambdas{1}{\delta^0, -\ofrac{l} \log \frac{\P_1(U=0)}{\P_0(U=0)}-\epsilon} + o(1),
\end{align*}
where $o(1)$ is a term that goes to zero as $n\to \infty$.
Taking $n\to\infty$ and then $\epsilon \to 0$, we obtain
\begin{align}
& \limsup_{n\to\infty} \ofrac{n} \log \P_1(Y_f=0\mid U = 0) + \limsup_{n\to\infty} \ofrac{n} \log \P_1(U = 0) \nonumber\\
& \geq - (1-r)\sup_{\delta^0\in\Gamma} \Lambdas{1}{\delta^0, \frac{r}{1-r}(e_{10} - e_{00})} - re_{10} \nonumber\\
& = - (1-r) \sup_{\delta^0\in\Gamma} \Lambdas{0}{\delta^0, \frac{r}{1-r}(e_{10} - e_{00})} - re_{00}\label{eq:TypeI0}
\end{align}
In the same way, it can be checked that
\begin{align}
& \limsup_{n\to\infty} \ofrac{n} \log \P_1(Y_f=0\mid U = 1) + \limsup_{n\to\infty} \ofrac{n} \log \P_1(U = 1) \nonumber\\
& \geq - (1-r) \sup_{\delta^1\in\Gamma} \Lambdas{1}{\delta^1, -\frac{r}{1-r}(e_{01} - e_{11})} - re_{11}, \label{eq:TypeI1}
\end{align}
and we obtain
\begin{align*}
\limsup_{n\to\infty} \ofrac{n} \log \P_1(Y_f=0) 
& \geq - h(\vec{e}). 
\end{align*}
A similar proof shows that 
\begin{align*}
\limsup_{n\to\infty} \ofrac{n} \log \P_0(Y_f=1) 
& \geq - h(\vec{e}),
\end{align*}
and that the optimal error exponent is lower bounded by $-h(\vec{e})$. We note that this lower bound can be asymptotically achieved by letting all sensors in the second stage quantize their observations using $\delta^0$ and $\delta^1$, depending on whether the feedback message is 0 or 1 respectively, and where $\delta^0$ and $\delta^1$ are chosen to asymptotically maximize their respective rate functions in \eqref{eq:h}.
The proof is now complete.
\end{IEEEproof}

Reference \cite{Zou:Msc09} shows that if $e_{01} > 0$ and $e_{10}>0$, then there is no loss in optimality if  sensors within each stage are constrained to using the same quantization function. In the following, we show that it is optimal to require that $e_{01} > 0$ and $e_{10}>0$. We also provide a characterization of the optimal error exponent.  

\begin{Theorem}\label{theorem:Bayesian-RestrictedFBDC}
Suppose that Assumptions \ref{assumpt:EquivalentMeasures} and \ref{assumpt:BoundedDivergence} hold. Then, the following statements hold for a RFDC architecture.
\begin{enumerate}[(i)]
\item\label{it:stage1} There is no loss in optimality if $e_{01}$ and $e_{10}$ are constrained to be strictly positive.
\item\label{it:quantizers1} There is no loss in optimality if sensors in the first stage are constrained to using the same quantization function.
\item\label{it:quantizers2} There is no loss in optimality if sensors in the second stage are constrained to using the same quantization function (which may depend on the feedback message).
\item\label{it:errexp} The optimal error exponent for the RFDC is
\begin{align}
\mathcal{E}^*_{dc} = -(1-r)\sup_{\substack{\gamma,\delta^0,\delta^1\in\Gamma \\ t\in\Real}}
\min\left\{\Lambdas{0}{\delta^0, \frac{r}{1-r} \Lambdas{1}{\gamma,t}}, 
\Lambdas{1}{\delta^1, -\frac{r}{1-r} \Lambdas{0}{\gamma,t}} \right\}. \label{eq:Edc}
\end{align}
\end{enumerate}
\end{Theorem}
\begin{IEEEproof}
We first show claim \eqref{it:stage1}. Note that only one of $e_{01}$ and $e_{00}$ can be strictly positive. The same applies to $e_{10}$ and $e_{11}$. If $e_{01} > 0$ and $e_{10} > 0$, we have $e_{00} = e_{11} = 0$, and \eqref{eq:h} yields
\begin{align*}
h(\vec{e}) 
& = (1-r) \min\Big\{\sup_{\delta\in\Gamma} \Lambdas{0}{\delta,\frac{r}{1-r}e_{10}},\sup_{\delta\in\Gamma} \Lambdas{1}{\delta,-\frac{r}{1-r}e_{01}} \Big\} \\
& \geq (1-r) \min\Big\{\sup_{\delta\in\Gamma} \Lambdas{0}{\delta,0},\sup_{\delta\in\Gamma} \Lambdas{1}{\delta,0} \Big\}\\
& = (1-r) \sup_{\delta\in\Gamma} \Lambdas{1}{\delta,0}.
\end{align*}
Suppose that $e_{00} > 0$ and $e_{10} > 0$. Then, $e_{01} = e_{11} = 0$, and from \eqref{eq:h}, we have $h(\vec{e}) \leq (1-r) \sup_{\delta\in\Gamma} \Lambdas{1}{\delta,0}$. The same argument applies for the case where $e_{01} > 0$ and $e_{11} > 0$, and the case where all the decay rates are zero. Therefore, there is no loss in optimality if $e_{01}$ and $e_{10}$ are constrained to be strictly positive.
 
Claims \eqref{it:quantizers1} and \eqref{it:quantizers2} follow from either an application of Cram\`er's Theorem (cf.\ \cite{Str:93}) and \eqref{eq:h}, or from \cite{Zou:Msc09}. 

Finally, we prove claim \eqref{it:errexp}. Since there is no loss in optimality if all first stage sensors are restricted to some same quantization function $\gamma\in\Gamma$, the first stage Type I and II error decay rates are $e_{01} = \Lambdas{0}{\gamma,t}$ and $e_{10} = \Lambdas{1}{\gamma,t}$ respectively, for some $t$ (cf.\ \cite{TayTsiWin:J09a}). Applying Lemma \ref{lemma:RestrictedFBDCLowerBound}, and optimizing over $\gamma$ and $t$, we have shown that the optimal error exponent is lower bounded by the right hand side of \eqref{eq:Edc}. This bound is achievable, hence the claim follows. The proof is now complete.
\end{IEEEproof}

Let $\mathcal{E}^*_{t}$ be the optimal error exponent of the daisy-chain if the second stage sensors ignore the feedback message. This is equivalent to a tree architecture with a height of two. Using the same arguments as above, it can be shown that
\begin{align}
\mathcal{E}^*_{t} = -(1-r)\sup_{\substack{\gamma,\delta\in\Gamma \\ t\in\Real}}
\min\left\{\Lambdas{0}{\delta, \frac{r}{1-r} \Lambdas{1}{\gamma,t}}, 
\Lambdas{1}{\delta, -\frac{r}{1-r} \Lambdas{0}{\gamma,t}} \right\}. \label{eq:Et}
\end{align}

Comparing \eqref{eq:Edc} and \eqref{eq:Et}, we have $\mathcal{E}^*_{dc} \leq \mathcal{E}^*_{t}$, i.e., the optimal error exponent for the RFDC is in general better than the tree configuration where feedback is absent. In the following, we provide a sufficient condition for no loss in performance when feedback is ignored, i.e., $\mathcal{E}^*_{dc} = \mathcal{E}^*_{t}$. We also provide a numerical example in which $\mathcal{E}^*_{dc} < \mathcal{E}^*_{t}$, i.e., feedback can strictly improve the asymptotic performance in some cases. 

\begin{Proposition}\label{prop:SymmetricRate}
Suppose that there exists $\delta \in \Gamma$ such that 
\begin{align}
\mathcal{E}^*_{t} = -(1-r)\sup_{\substack{\gamma\in\Gamma \\ t\in\Real}}
\min\left\{\Lambdas{0}{\delta, \frac{r}{1-r} \Lambdas{1}{\gamma,t}}, 
\Lambdas{1}{\delta, -\frac{r}{1-r} \Lambdas{0}{\gamma,t}} \right\}, \label{eq:Et_exp}
\end{align}
and $\Lambdas{1}{\delta, t} = \Lambdas{0}{\delta, -t}$ for all $t$. Then,
\begin{align*}
\mathcal{E}^*_{dc} = \mathcal{E}^*_{t} = -(1-r) \sup_{\gamma,\delta\in\Gamma} 
\Lambdas{0}{\delta, \frac{r}{1-r} \Lambdas{1}{\gamma,0}}.
\end{align*}
Therefore, there is no loss in optimality if the RFDC second stage sensors ignore the feedback message. 
\end{Proposition}
\begin{IEEEproof}
To simplify the proof, we assume that $\gamma \in \Gamma$ and $t\in\Real$ can be chosen so that the supremum in \eqref{eq:Et_exp} is achieved. To find the optimal 
threshold $t$, we set 
\begin{align*}
\Lambdas{0}{\delta, \frac{r}{1-r} \Lambdas{1}{\gamma,t}} = \Lambdas{1}{\delta, -\frac{r}{1-r} \Lambdas{0}{\gamma,t}}.
\end{align*}
From the proposition hypothesis, we obtain
$\Lambdas{1}{\gamma, t} = \Lambdas{0}{\gamma,t}$, which implies that $t=0$.
Therefore, 
\begin{align}\label{Et_sym}
\mathcal{E}^*_{t} = -(1-r) \sup_{\gamma,\delta\in\Gamma} \Lambdas{0}{\delta, \frac{r}{1-r} \Lambdas{1}{\gamma,0}}.
\end{align}

Suppose that there exists $\delta^0 \neq \delta^1$, and $v \neq 0$ such that 
\begin{align}
\min\left\{\Lambdas{0}{\delta^0, \frac{r}{1-r} \Lambdas{1}{\gamma,v}}, 
\Lambdas{1}{\delta^1, -\frac{r}{1-r} \Lambdas{0}{\gamma,v}} \right\} > \Lambdas{0}{\delta, \frac{r}{1-r} \Lambdas{1}{\gamma,0}}. \label{ineq:grate}
\end{align}
If $v > 0$, $\Lambdas{1}{\gamma,v} \leq \Lambdas{1}{\gamma,0}$ since $\Lambdas{1}{\gamma,\cdot}$ is a decreasing function. Therefore, 
\begin{align*}
\Lambdas{0}{\delta^0, \frac{r}{1-r} \Lambdas{1}{\gamma,0}} 
& \geq \Lambdas{0}{\delta^0, \frac{r}{1-r} \Lambdas{1}{\gamma,v}} > \Lambdas{0}{\delta, \frac{r}{1-r} \Lambdas{1}{\gamma,0}},
\end{align*}
a contradiction to \eqref{Et_sym}. A similar argument produces a contradiction if $v < 0$. Therefore, we must have $v=0$. But this implies that \eqref{ineq:grate} cannot hold. Hence, $\mathcal{E}^*_{dc} = \mathcal{E}^*_{t}$, and the proposition is proved.
\end{IEEEproof}

The following example shows that in some cases, the RFDC performs strictly better in the presence of feedback. 
\begin{Example}\label{ex:RFDC}
Let $X_k$ take values in the set $\{1,2,3\}$, and suppose that sensor messages are restricted to a single bit. Assume that the probability mass functions under the two hypotheses are as shown in Table \ref{table:pmf}. We also let $m = n/2$, i.e., $r = 1-r = 1/2$. 
\begin{table}[!ht]
\begin{center}
\begin{tabular}{|c|c|c|c|}
\hline
       & 1 & 2 & 3 \\ \hline
$H_0$  & 4/5 & 3/20 & 1/20 \\ \hline
$H_1$  & 1/20 & 3/20 & 4/5 \\ \hline 
\end{tabular}\\
\ \\
\caption{Probability mass functions for Example \ref{ex:RFDC}.}\label{table:pmf}
\end{center}
\end{table}

Since $\Lh{10}(X_k)$ is increasing with $X_k$, the two possible 1-bit quantizers are $\gamma_1(X_k) = 0$ iff $X_k = 1$, and $\gamma_2(X_k) = 0$ iff $X_k \in \{1,2\}$. We optimize \eqref{eq:Edc} over these two quantizers and the threshold $t$. The results are shown in Figure \ref{fig:optresults}. The optimal error exponent is found to be $-0.5\cdot 0.73 = -0.365$, and is achieved by having all second stage sensors use $\gamma_2$ if the feedback message is 0, and $\gamma_1$ if the feedback message is 1. On the other hand, if feedback is ignored, the optimal quantizer is $\gamma_2$, and the optimal error exponent is $-0.356$, which is strictly worse than that with feedback.  

\begin{figure}[!htb]
\begin{center}
\includegraphics[height=3in]{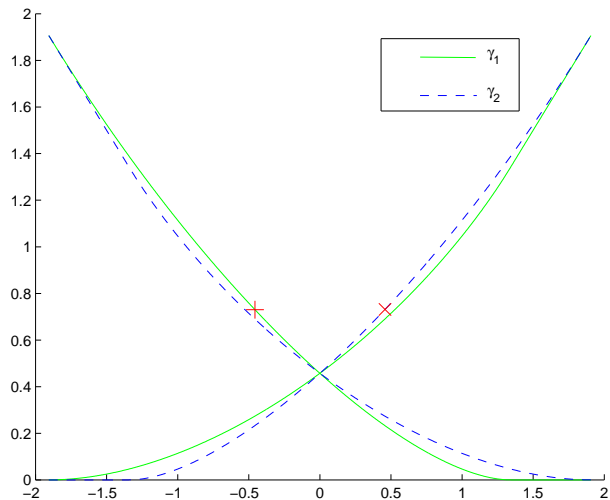}
\caption{Plot of the rate functions for $\gamma_1$ and $\gamma_2$. The mark 'x' indicates the optimal error decay rate (up to a constant $1/2$) when the feedback message $U=0$, while '+' indicates the optimal error decay rate (up to a constant $1/2$) when the feedback message $U=1$. The optimal quantizers are achieved on rate functions belonging to different quantizers.
}\label{fig:optresults}
\end{center}
\end{figure}
\end{Example}

In the following, we show that the RFDC performs strictly worse than a parallel configuration, and hence it has performance strictly inferior to a full feedback daisy-chain architecture. 
\begin{Proposition}\label{prop:DCvsParallel}
Suppose that the supremum in \eqref{eq:Edc} is achieved. 
Then, the RFDC performs strictly worse than the 
parallel configuration with the same total number of sensors, i.e.,
$\mathcal{E}^*_{t} \geq \mathcal{E}^*_{dc} > \mathcal{E}^*_{1p} = -\sup_{\delta\in\Gamma} \Lambdas{0}{\delta, 0}$.
\end{Proposition}
\begin{IEEEproof}
Let $\gamma,\delta^0,\delta^1,t$ achieve the supremum in \eqref{eq:Edc}.
If $\Lambdas{1}{\gamma,t} = 0$, then from \eqref{eq:Edc}, we have
\begin{align*}
\mathcal{E}^*_{dc} 
& \geq -(1-r) \Lambdas{0}{\delta^0, 0} \\
& > - \sup_{\delta\in\Gamma} \Lambdas{0}{\delta, 0},
\end{align*}
since $r > 0$. 
A similar argument shows that $\mathcal{E}^*_{dc} > \mathcal{E}^*_{1p}$ if $\Lambdas{0}{\gamma,t} = 0$. Therefore, in the following, we assume that $\Lambdas{j}{\gamma,t} > 0$ for $j=0,1$.  

We have
\begin{align}
& (1-r)\Lambdas{0}{\delta^0, \frac{r}{1-r} \Lambdas{1}{\gamma,t}} \nonumber\\
& = (1-r)\Lambdas{1}{\delta^0, \frac{r}{1-r} \Lambdas{1}{\gamma,t}} + r\Lambdas{1}{\gamma,t} \nonumber\\
& < (1-r) \Lambdas{1}{\delta^0,0} + r\Lambdas{1}{\gamma,t} \nonumber\\
& \leq (1-r) \sup_{\delta\in\Gamma} \Lambdas{1}{\delta,0} + r\Lambdas{1}{\gamma,t},\label{eq:Lambdas0Bound}
\end{align}
where the penultimate inequality follows from $\Lambdas{1}{\delta^0,\cdot}$ being a decreasing 
function, and $\Lambdas{1}{\gamma,t} > 0$. 
Similarly,
\begin{align}
& (1-r)\Lambdas{1}{\delta^1, -\frac{r}{1-r} \Lambdas{0}{\gamma,t}} \nonumber\\
& < (1-r) \sup_{\delta\in\Gamma} \Lambdas{0}{\delta,0} + r\Lambdas{0}{\gamma,t}.\label{eq:Lambdas1Bound}
\end{align}
Combining \eqref{eq:Lambdas0Bound} and \eqref{eq:Lambdas1Bound}, and since $\Lambdas{0}{\delta,0} = \Lambdas{1}{\delta,0}$ for all $\delta\in\Gamma$, we obtain
\begin{align*}
\mathcal{E}^*_{dc}
& > -(1-r) \sup_{\delta\in\Gamma} \Lambdas{0}{\delta,0} - r \min\left\{\Lambdas{1}{\gamma,t},\Lambdas{0}{\gamma,t}\right\} \\
& \geq -(1-r) \sup_{\delta\in\Gamma} \Lambdas{0}{\delta,0} - r \Lambdas{0}{\gamma,0} \\
& \geq - \sup_{\delta\in\Gamma} \Lambdas{0}{\delta,0} = \mathcal{E}^*_{1p}.
\end{align*}
The proof is now complete.

\end{IEEEproof}

\section{Conclusion}\label{sect:Conclusion}

We have studied two-message feedback architectures, in which each sensor has access to compressed summaries of some or all other sensors' first messages to the fusion center. In the sequential feedback architecture, each sensor has access to the first messages of those sensors that communicate with the fusion center before it. In the restricted and full feedback architectures, each sensor has partial and full information respectively, about the first messages of every other sensor. Under both the Neyman-Pearson and Bayesian formulations, we show that the optimal error exponent is not improved by the feedback messages. We have also studied the one-message feedback architectures in which a group of sensors have access to information from sensors in a first group. We show that if the fusion center has knowledge of all the messages from the sensors in the first group, then feedback does not improve the optimal error exponent, which is the same as the parallel configuration. In the case where the fusion center has only limited knowledge (a 1-bit summary) of the messages, feedback can improve the optimal error exponent, but the optimal error exponent is strictly worse than that of the parallel configuration. Our results suggest that in the regime of a large number of sensors, and where the fusion center has sufficient memory, the performance gain in binary hypothesis testing due to feedback does not justify the increase in communication and computation costs incurred in a feedback architecture. 

In the two-message feedback architecture, we assumed that the fusion center has unlimited memory and remembers all the first messages. The case where the fusion center retains only a finitely-valued summary of the first messages has been studied in \cite{ShaPap:93}, but under various assumptions including finitely-value observation spaces, sensors all using the same quantization functions and constraints on the feedback messages. Reference \cite{ShaPap:93} shows that feedback does not improve the error exponent. The same problem in the general setting that we have considered in this paper remains open.

In the case of Bayesian $M$-ary hypothesis testing, where $M > 2$, we conjecture that feedback improves the optimal error exponent. Characterizing the optimal feedback strategy and error exponent is part of future work. This research is also part of our ongoing efforts to quantify the performance of various network architectures. Future research directions include studying network architectures with more general loop structures.

\section{Acknowledgements}\label{sect:Acknowledgements}

The author wishes to thank Professor John Tsitsiklis for fruitful discussions and numerous suggestions that led to simplifications of some proofs, and significant improvements in the organization and presentation of this paper.

\appendices

\section{Mathematical Preliminaries}\label{appendix:Mathematical}

In this appendix, we collect two well known results that are useful in our proofs.
The first result is an elementary fact, which is an application of Jensen's inequality. A proof can be found in \cite{Tsi:88}, and is omitted here.
\begin{Proposition_A}\label{proposition:JensenDiv}
Suppose $\phi : (0,\infty) \mapsto \Real$ is a convex function. 
Then for any function $\gamma$, we have
\begin{align*}
\E_j\left[\phi\left(\Lh{ij}(\gamma(X))\right)\right] \leq \E_j\left[\phi\left(\Lh{ij}(X)\right)\right].
\end{align*}
\end{Proposition_A}

The following lower bound for the maximum of the Type I and II error probabilities was first proved in \cite{ShaGalBer:67a} for the case of discrete observation spaces. The following proposition generalizes the result to a general observation space. The proof is identical to that in \cite{ShaGalBer:67a}, with some notation changes, and is provided for completeness.

\begin{Proposition_A}\label{proposition:errorproblowerbound}
Consider a hypothesis testing problem based on an observation $X$ with distribution $\P_j$ under hypothesis $H_j$, $j=0,1$. Suppose that the measures $\P_0$ and $\P_1$ are absolutely continuous w.r.t.\ each other. Let $P_{e,j}$ be the probability of error when $H_j$ is true.
Let $Z = \log \ddfrac{\P_1}{\P_0}(X)$ be the log Radon-Nikodym derivative.
For any $s \in \Real$, let $\Lambda(s) = \log \E_0[\exp(s Z)]$ be the log-moment generating function of $Z$. Then, for $s^*\in [0,1]$ such that $\Lambda'(s^*) = 0$, we have 
\begin{align*}
\max(P_{e,0},P_{e,1}) \geq \ofrac{4} \exp\left(\Lambda(s^*) - \sqrt{2\Lambda''(s^*)}\right).
\end{align*}
\end{Proposition_A}
\begin{proof}
The proof steps are identical to that of Theorem 5 in \cite{ShaGalBer:67a}. 
Let $P_j$ be the probability measure of $Z$ under hypothesis $H_j$, $j=0,1$.
For $s \in (0,1)$, define the probability measure $Q$ such that 
\begin{align*}
\ddfrac{Q}{P_0}(z) = e^{sz - \Lambda(s)},
\end{align*}
and let $\E_Q$ and $\var_Q$ be the mathematical expectation and variance w.r.t.\ $Q$, respectively. 
Let $Y$ be a random variable with distribution $Q$. 
Then, it is easy to check that $\E_Q[Y] = \Lambda'(s)$ and 
$\var_Q(Y) = \Lambda''(s)$. 
Let $A_s = \{y : |y - \Lambda'(s)| \leq \sqrt{2\Lambda''(s)}\}$. From Chebychev's inequality, we have
\begin{align}\label{nuAs}
Q(A_s) > \ofrac{2}.
\end{align}
For any measurable set $A$, we have
\begin{align*}
P_0(A) & = \E_Q[\exp(-sZ + \Lambda(s))\indicator{Z \in A}]\\
& \geq  \E_Q[\exp(-sZ + \Lambda(s))\indicator{Z \in A\cap A_s}]\\
& \geq \exp\big(\Lambda(s) - s\Lambda'(s) - s\sqrt{2\Lambda''(s)}\big) \ Q(A\cap A_s).
\end{align*}
Similarly, we have
\begin{align*}
P_1(A^c) 
& \geq \exp\big(\Lambda(s)+(1-s)\Lambda'(s) - (1-s)\sqrt{2\Lambda''(s)}\big) \ Q(A^c\cap A_s).
\end{align*}
From \eqref{nuAs}, either $Q(A\cap A_s) > 1/4$ or $Q(A^c\cap A_s) > 1/4$. Therefore, we have either 
\begin{align}\label{ineq:z0}
P_0(A)
& \geq \ofrac{4}\exp\big(\Lambda(s) - s\Lambda'(s) - s\sqrt{2\Lambda''(s)}\big),
\end{align}
or
\begin{align}\label{ineq:z1}
P_1(A^c)
& \geq \ofrac{4}\exp\big(\Lambda(s)+(1-s)\Lambda'(s) - (1-s)\sqrt{2\Lambda''(s)}\big).
\end{align}
Since $\Lambda(s)$ is convex with $\Lambda(0) = \Lambda(1) = 0$, there exists $s^* \in (0,1)$ such that 
$\Lambda(s^*) = 0$. Substituting this into \eqref{ineq:z0} and \eqref{ineq:z1}, we obtain
\begin{align*}
\max(P_0(A), P_1(A^c)) 
& \geq \ofrac{4}\exp\big(\Lambda(s^*) - \sqrt{2\Lambda''(s^*)}\big).
\end{align*}
The proof is now complete. 
\end{proof}

\bibliographystyle{IEEEtran}
\bibliography{IEEEabrv,StringDefinitions,DecentDet,BibBooks,Tay}

\begin{thebibliography}{10}
\providecommand{\url}[1]{#1}
\csname url@samestyle\endcsname
\providecommand{\newblock}{\relax}
\providecommand{\bibinfo}[2]{#2}
\providecommand{\BIBentrySTDinterwordspacing}{\spaceskip=0pt\relax}
\providecommand{\BIBentryALTinterwordstretchfactor}{4}
\providecommand{\BIBentryALTinterwordspacing}{\spaceskip=\fontdimen2\font plus
\BIBentryALTinterwordstretchfactor\fontdimen3\font minus
  \fontdimen4\font\relax}
\providecommand{\BIBforeignlanguage}[2]{{%
\expandafter\ifx\csname l@#1\endcsname\relax
\typeout{** WARNING: IEEEtran.bst: No hyphenation pattern has been}%
\typeout{** loaded for the language `#1'. Using the pattern for}%
\typeout{** the default language instead.}%
\else
\language=\csname l@#1\endcsname
\fi
#2}}
\providecommand{\BIBdecl}{\relax}
\BIBdecl

\bibitem{TenSan:81}
R.~R. Tenney and N.~R. Sandell, ``Detection with distributed sensors,''
  \emph{{IEEE} Trans. Aerosp. Electron. Syst.}, vol.~17, no.~4, pp. 501--510,
  1981.

\bibitem{ChaVar:86}
Z.~Chair and P.~K. Varshney, ``Optimal data fusion in multiple sensor detection
  systems,'' \emph{{IEEE} Trans. Aerosp. Electron. Syst.}, vol.~22, no.~1, pp.
  98--101, 1986.

\bibitem{PolTsi:90}
G.~Polychronopoulos and J.~N. Tsitsiklis, ``Explicit solutions for some simple
  decentralized detection problems,'' \emph{{IEEE} Trans. Aerosp. Electron.
  Syst.}, vol.~26, no.~2, pp. 282--292, 1990.

\bibitem{WilWar:92}
P.~Willett and D.~Warren, ``The suboptimality of randomized tests in
  distributed and quantized detection systems,'' \emph{{IEEE} Trans. Inf.
  Theory}, vol.~38, no.~2, pp. 355--361, Mar. 1992.

\bibitem{Tsi:93a}
J.~N. Tsitsiklis, ``Extremal properties of likelihood-ratio quantizers,''
  \emph{{IEEE} Trans. Commun.}, vol.~41, no.~4, pp. 550--558, 1993.

\bibitem{Tsi:93}
------, ``Decentralized detection,'' \emph{Advances in Statistical Signal
  Processing}, vol.~2, pp. 297--344, 1993.

\bibitem{IrvTsi:94}
W.~W. Irving and J.~N. Tsitsiklis, ``Some properties of optimal thresholds in
  decentralized detection,'' \emph{{IEEE} Trans. Autom. Control}, vol.~39,
  no.~4, pp. 835--838, 1994.

\bibitem{VisVar:97}
R.~Viswanathan and P.~K. Varshney, ``Distributed detection with multiple
  sensors: part {I} - fundamentals,'' \emph{Proc. {IEEE}}, vol.~85, no.~1, pp.
  54--63, 1997.

\bibitem{CheVar:02}
B.~Chen and P.~K. Varshney, ``A {Bayesian} sampling approach to decision fusion
  using hierarchical models,'' \emph{{IEEE} Trans. Signal Process.}, vol.~50,
  no.~8, pp. 1809--1818, Aug. 2002.

\bibitem{ChaVee:03}
J.-F. Chamberland and V.~V. Veeravalli, ``Decentralized detection in sensor
  networks,'' \emph{{IEEE} Trans. Signal Process.}, vol.~51, no.~2, pp.
  407--416, 2003.

\bibitem{CheWil:05}
B.~Chen and P.~K. Willett, ``On the optimality of the likelihood-ratio test for
  local sensor decision rules in the presence of nonideal channels,''
  \emph{{IEEE} Trans. Inf. Theory}, vol.~51, no.~2, pp. 693--699, Feb. 2005.

\bibitem{Kas:06}
A.~Kashyap, ``Comments on ``{On} the optimality of the likelihood-ratio test
  for local sensor decision rules in the presence of nonideal channels'',''
  \emph{{IEEE} Trans. Inf. Theory}, vol.~52, no.~3, pp. 1274--1275, Mar. 2006.

\bibitem{LiuChe:06}
B.~Liu and B.~Chen, ``Channel-optimized quantizers for decentralized detection
  in sensor networks,'' \emph{{IEEE} Trans. Inf. Theory}, vol.~52, no.~7, pp.
  3349--3358, Jul. 2006.

\bibitem{VisThoTum:88}
R.~Viswanathan, S.~C.~A. Thomopoulos, and R.~Tumuluri, ``Optimal serial
  distributed decision fusion,'' \emph{{IEEE} Trans. Aerosp. Electron. Syst.},
  vol.~24, no.~4, pp. 366--376, 1988.

\bibitem{TanPatKle:91}
Z.~B. Tang, K.~R. Pattipati, and D.~L. Kleinman, ``Optimization of detection
  networks: part {I} - tandem structures,'' \emph{{IEEE} Trans. Syst., Man,
  Cybern.}, vol.~21, no.~5, pp. 1044--1059, 1991.

\bibitem{PapAth:92}
J.~D. Papastavrou and M.~Athans, ``Distributed detection by a large team of
  sensors in tandem,'' \emph{{IEEE} Trans. Aerosp. Electron. Syst.}, vol.~28,
  no.~3, pp. 639--653, 1992.

\bibitem{TayTsiWin:J08c}
W.~P. Tay, J.~N. Tsitsiklis, and M.~Z. Win, ``On the sub-exponential decay of
  detection error probabilities in long tandems,'' \emph{{IEEE} Trans. Inf.
  Theory}, vol.~54, no.~10, pp. 4767--4771, Oct. 2008.

\bibitem{EkcTen:82}
L.~K. Ekchian and R.~R. Tenney, ``Detection networks,'' in \emph{Proc. IEEE
  Conf. Decision and Control}, vol.~21, Dec. 1982, pp. 686--691.

\bibitem{ReiNol:90a}
A.~R. Reibman and L.~W. Nolte, ``Design and performance comparison of
  distributed detection networks,'' \emph{{IEEE} Trans. Aerosp. Electron.
  Syst.}, vol.~23, no.~6, pp. 789--797, 1987.

\bibitem{TanPatKle:93}
Z.~B. Tang, K.~R. Pattipati, and D.~L. Kleinman, ``Optimization of detection
  networks: part {II} - tree structures,'' \emph{{IEEE} Trans. Syst., Man,
  Cybern.}, vol.~23, no.~1, pp. 211--221, 1993.

\bibitem{PetPatKle:94}
A.~Pete, K.~Pattipati, and D.~Kleinman, ``Optimization of detection networks
  with multiple event structures,'' \emph{{IEEE} Trans. Autom. Control},
  vol.~39, no.~8, pp. 1702--1707, 1994.

\bibitem{AlhVar:95}
S.~Alhakeem and P.~K. Varshney, ``A unified approach to the design of
  decentralized detection systems,'' \emph{{IEEE} Trans. Aerosp. Electron.
  Syst.}, vol.~31, no.~1, pp. 9--20, 1995.

\bibitem{LinCheVar:05}
Y.~Lin, B.~Chen, and P.~K. Varshney, ``Decision fusion rules in multi-hop
  wireless sensor networks,'' \emph{{IEEE} Trans. Aerosp. Electron. Syst.},
  vol.~41, no.~2, pp. 475--488, Apr. 2005.

\bibitem{TayTsiWin:J08a}
W.~P. Tay, J.~N. Tsitsiklis, and M.~Z. Win, ``On the impact of node failures
  and unreliable communications in dense sensor networks,'' \emph{{IEEE} Trans.
  Signal Process.}, vol.~56, no.~6, pp. 2535--2546, Jun. 2008.

\bibitem{TayTsiWin:J08b}
------, ``Data fusion trees for detection: Does architecture matter?''
  \emph{{IEEE} Trans. Inf. Theory}, vol.~54, no.~9, pp. 4155--4168, Sep. 2008.

\bibitem{TayTsiWin:J09a}
------, ``Bayesian detection in bounded height tree networks,'' \emph{{IEEE}
  Trans. Signal Process.}, vol.~57, no.~10, pp. 4042--4051, Oct 2009.

\bibitem{TsiAth:85}
J.~N. Tsitsiklis and M.~Athans, ``On the complexity of decentralized decision
  making and detection problems,'' \emph{{IEEE} Trans. Autom. Control},
  vol.~30, pp. 440--446, 1985.

\bibitem{ChaVee:06}
J.-F. Chamberland and V.~V. Veeravalli, ``How dense should a sensor network be
  for detection with correlated observations?'' \emph{{IEEE} Trans. Inf.
  Theory}, vol.~52, no.~11, pp. 5099--5106, Nov. 2006.

\bibitem{MisTon:08}
S.~Misra and L.~Tong, ``Error exponents for the detection of {Gauss-Markov}
  signals using randomly spaced sensors,'' \emph{Signal Processing, IEEE
  Transactions on}, vol.~56, no.~8, pp. 3385 --3396, Aug 2008.

\bibitem{SunPooYu:09}
Y.~Sung, H.~V. Poor, and H.~Yu, ``How much information can one get from a
  wireless ad hoc sensor network over a correlated random field?'' \emph{{IEEE}
  Trans. Inf. Theory}, vol.~55, no.~6, pp. 2827--2847, Jun 2009.

\bibitem{KreWil:10}
O.~Kreidl and A.~Willsky, ``An efficient message-passing algorithm for
  optimizing decentralized detection networks,'' \emph{{IEEE} Trans. Autom.
  Control}, vol.~55, no.~3, pp. 563--578, Mar 2010.

\bibitem{AlhVar:96}
S.~Alhakeem and P.~Varshney, ``Decentralized {Bayesian} detection with
  feedback,'' \emph{{IEEE} Trans. Syst., Man, Cybern.}, vol.~26, no.~4, pp.
  503--513, Jul. 1996.

\bibitem{ShaPap:93}
H.~Shalaby and A.~Papamarcou, ``A note on the asymptotics of distributed
  detection with feedback,'' \emph{Information Theory, IEEE Transactions on},
  vol.~39, no.~2, pp. 633--640, Mar 1993.

\bibitem{Zou:Msc09}
S.~I. Zoumpoulis, ``Decentralized detection in sensor network architectures
  with feedback,'' Master's thesis, Massachusetts Institute of Technology, June
  2009.

\bibitem{KreTsiZou:09}
O.~P. Kreidl, J.~N. Tsitsiklis, and S.~I. Zoumpoulis, ``On decentralized
  detection with partial information sharing among sensors,'' \emph{{IEEE}
  Trans. Signal Process.}, 2011, in press.

\bibitem{SmiSor:00}
L.~Smith and P.~Sorensen, ``Pathological outcomes of observational learning,''
  \emph{Econometrica}, vol.~68, pp. 371--398, 2000.

\bibitem{AceDahLobOzd:11}
D.~Acemoglu, M.~A. Dahleh, I.~Lobel, and A.~Ozdaglar, ``Bayesian learning in
  social networks,'' \emph{Review of Economic Studies}, 2011, forthcoming.

\bibitem{TayTsiWin:J07a}
W.~P. Tay, J.~N. Tsitsiklis, and M.~Z. Win, ``Asymptotic performance of a
  censoring sensor network,'' \emph{{IEEE} Trans. Inf. Theory}, vol.~53,
  no.~11, pp. 4191--4209, Nov. 2007.

\bibitem{Tsi:88}
J.~N. Tsitsiklis, ``Decentralized detection by a large number of sensors,''
  \emph{Math. Control, Signals, Syst.}, vol.~1, pp. 167--182, 1988.

\bibitem{DemZei:98}
A.~Dembo and O.~Zeitouni, \emph{Large Deviations Techniques and
  Applications}.\hskip 1em plus 0.5em minus 0.4em\relax New York, NY:
  Springer-Verlag, 1998.

\bibitem{LehRom:B05}
E.~Lehmann and J.~P. Romano, \emph{Testing Statistical Hypotheses}.\hskip 1em
  plus 0.5em minus 0.4em\relax New York, NY: Springer, 2005.

\bibitem{Str:93}
D.~Stroock, \emph{Probability Theory: An Analytic View}.\hskip 1em plus 0.5em
  minus 0.4em\relax Cambridge, UK: Cambridge University Press, 1993.

\bibitem{ShaGalBer:67a}
C.~E. Shannon, R.~G. Gallager, and E.~R. Berlekamp, ``Lower bounds to error
  probability for coding on discrete memoryless channels, {I},''
  \emph{Information and Control}, vol.~10, pp. 65--103, 1967.

\end{thebibliography}

\end{document}